\documentclass{aa}

\usepackage{txfonts}
\usepackage{longtable}
\usepackage{natbib}
\usepackage{graphicx}

\bibpunct[ ]{(}{)}{;}{a}{}{,}  % LM
\usepackage{supertabular}
\usepackage{color}
\newcommand{\eps}[1]{\log\varepsilon_{\rm #1}}
\newcommand{\kms}{km\,s$^{-1}$}
\newcommand{\Vmic}{$V_{\rm mic}$}
\newcommand{\Vmac}{V_{\rm mac}}
\newcommand{\eu}[5]{\mbox{$#1\,^#2{\rm #3}^{#4}_{\rm #5}$}}
\newcommand{\Teff}{T_{\rm eff}}
\newcommand{\Eexc}{$E_{\rm exc}$}
\newcommand{\kH}{$S_{\!\!\rm H}$}    %%% Note negative spaces!
\newcommand{\logg}{\rm log~ g}

\begin{document}

\title{A non-LTE study of neutral and singly-ionized iron line \\ spectra in 1D models of the Sun and selected late-type stars \thanks{Based on observations collected at the German Spanish Astronomical Center, Calar Alto, Spain and taken from the ESO UVES-POP archive}}

\author{
 L. Mashonkina\inst{1,2} \and
 T. Gehren\inst{1}  \and
 J.-R. Shi\inst{3}  \and
 A. J. Korn\inst{4}  \and
 F. Grupp\inst{1,5}
}
\offprints{L. Mashonkina; \email{lyuda@usm.lmu.de}}
\institute{
     Universit\"ats-Sternwarte M\"unchen, Scheinerstr. 1, D-81679 M\"unchen,
     Germany \\ \email{lyuda@usm.lmu.de}
\and Institute of Astronomy, Russian Academy of Sciences, RU-119017 Moscow,
     Russia \\ \email{lima@inasan.ru}
\and National Astronomical Observatories, Chinese Academy of Sciences, A20 Datun Road, Chaoyang District, Beijing 100012, PR China
\and Department of  Physics and Astronomy, Division of Astronomy and Space Physics, Uppsala University, Box 516, 75120 Uppsala, Sweden
\and Max-Planck Institut f\"ur Extraterrestrische Physik, Giessenbachstr.,
D-85748 Garching, Germany}

\date{Received  / Accepted }

\abstract
{}{Based on a rather complete model atom for neutral and singly-ionized iron, we evaluate non-local thermodynamical equilibrium (non-LTE) line formation for the two ions of iron and check the ionization equilibrium between \ion{Fe}{i} and \ion{Fe}{ii} in model atmospheres of the cool reference stars.}
{A comprehensive model atom for Fe with more than 3\,000 measured and predicted energy levels is presented. As a test and first application of the improved model atom, iron abundances are determined for the Sun and five stars with well determined stellar parameters and high-quality observed spectra. The efficiency of inelastic collisions with hydrogen atoms in the statistical equilibrium of iron is estimated empirically from inspection of their different influence on the \ion{Fe}{i} and \ion{Fe}{ii} lines in the selected stars.}
{Non-LTE leads to systematically depleted total absorption in the \ion{Fe}{i} lines and to positive abundance corrections in agreement with the previous studies, however, the magnitude of such corrections is smaller compared to the earlier results. Non-LTE corrections do not exceed 0.1~dex for the solar metallicity and mildly metal-deficient stars, and they vary within 0.21~dex and 0.35~dex in the very metal-poor stars HD~84937 and HD~122563, respectively, depending on the assumed efficiency of collisions with hydrogen atoms. Based on the analysis of the  \ion{Fe}{i}/\ion{Fe}{ii} ionization equilibrium in these two stars, we recommend to apply the Drawin formalism in non-LTE studies of Fe with a scaling factor of 0.1. For the \ion{Fe}{ii} lines, non-LTE corrections do not exceed 0.01~dex in absolute value. 
This study reveals two problems. The first one is that $gf-$values available for the \ion{Fe}{i} and \ion{Fe}{ii} lines are not accurate enough to pursue high-accuracy absolute stellar abundance determinations. For the Sun, the mean non-LTE abundance obtained from 54 \ion{Fe}{i} lines is 7.56$\pm$0.09 and the mean abundance from 18 \ion{Fe}{ii} lines varies between 7.41$\pm$0.11 and 7.56$\pm$0.05 depending on the source of the $gf-$values. The second problem is that lines of \ion{Fe}{i} give, on average, a 0.1\,dex lower abundance compared to those of \ion{Fe}{ii} lines for HD\,61421 and HD\,102870, 
even when applying a differential line-by-line analysis relative to the Sun. A disparity between neutral atoms and first ions points to problems of stellar atmosphere modelling or/and effective temperature determination.}
{}


\titlerunning{A non-LTE study of neutral and singly-ionized iron line spectra }
\authorrunning{Mashonkina et al.}

\maketitle

\section{Introduction}

Iron plays an outstanding role in studies of cool stars thanks to the many lines in the visible spectrum, which are easy to detect even in very
metal-poor stars. Iron serves as a reference element for all astronomical research related to stellar nucleosynthesis and the chemical evolution of the
Galaxy. Iron lines are used to derive basic stellar parameters, i.e. the effective temperature, $\Teff$, from the excitation equilibrium of \ion{Fe}{i} and the surface gravity, $\logg$, from the ionization equilibrium between \ion{Fe}{i} and \ion{Fe}{ii}. In stellar atmospheres with $\Teff > 4500$~K, neutral iron is a minority species, and its statistical equilibrium (SE) can easily deviate from thermodynamic equilibrium due to deviations of the mean intensity of ionizing radiation from the Planck function.
Therefore, since the beginning of the 1970s, a large number of studies attacked the problem of non-local thermodynamic
equilibrium (non-LTE) line formation for iron in the atmospheres of the Sun and cool stars. The original model atoms were from \citet{Tanaka71,Athay1972,Boyarchuketal85,Gigas86,Takeda91,Grattonetal99,Thevenin1999,Gehren2001a,ShTB01}, and \citet{Colletal05}, and they were widely applied in stellar parameter and abundance analyses \citep[see][for references]{aspl05}.
It was understood that the main non-LTE mechanism for \ion{Fe}{i} is ultra-violet (UV) overionization of the levels with excitation energy of 1.4 to 4.5~eV. This results in an underpopulation of neutral iron where all \ion{Fe}{i} lines are weaker than their LTE strengths, and it leads to positive non-LTE abundance corrections.

The need for a new analysis was motivated by the following problems uncovered by the previous non-LTE calculations for iron.

First, the results obtained for the populations of high-excitation levels of \ion{Fe}{i} were not always convincing. The highest levels presented in the model atom did not couple thermally to the ground state of \ion{Fe}{ii} indicating substantial term incompleteness. To force the levels near the continuum into LTE, an upper level thermalization procedure was applied \citep[more details in][]{Gehren2001a,Korn03,Colletal05}.

A second aspect is
the treatment of poorly known inelastic collisions with hydrogen atoms.
%In the atmosphere of stars close to solar metallicity, neutral H atoms typically outnumber electrons by a factor of 10$^4$, and by even larger factors in metal-poor stars.
Their role
in establishing the statistical equilibrium of atoms in cool stars is debated for decades, from \citet{gehren75} to \citet{barklem2010}. Experimental data on \ion{H}{i} collision cross-sections
are only available for the resonance transition in \ion{Na}{i} \citep{fleck91,belyaev99}, and detailed quantum mechanical calculations were published for the transitions between first nine levels in \ion{Li}{i} \citep{belyaev03,barklem2003} and \ion{Na}{i} \citep{belyaev99,belyaev2010,barklem2010}. For all other chemical species, the basic formula used to calculate collisions with \ion{H}{i} atoms is the one proposed by \citet{D68,D69}, as described by \citet{Steenbock1984}, and it suggests that their influence is comparable to electron impacts.
 The laboratory measurements and quantum mechanical calculations indicate that the Drawin formula overestimates rate coefficients for optically allowed transitions by one to seven orders of magnitude. Therefore, various approaches were employed in the literature to constrain empirically the efficiency of \ion{H}{i} collisions.
The studies of stellar \ion{Na}{i} lines favor a low efficiency of this type of collisions. For example,
\citet{h_cool_na_o} found that the center-to-limb variation of the solar \ion{Na}{i} 6160\AA\ line is reproduced in the non-LTE calculations  with pure electron collisions.
\citet{Grattonetal99} calibrated \ion{H}{i} collisions with sodium using RR~Lyr variables and concluded that the Drawin rates should be decreased by two orders of magnitude. Based on their solar \ion{Na}{i} line profile analysis, \citet{mg_c6} and \citet{Takeda95} recommended to scale the Drawin rates by a factor \kH\ = 0.05 and 0.1, respectively. With a similar value of \kH\ = 0.1, the ionization equilibrium between \ion{Ca}{i} and \ion{Ca}{ii} in selected metal-poor stars was matched consistently with surface gravities derived from {\sc Hipparcos} parallaxes \citep{mash_ca}. On the other hand, spectroscopic studies of different chemical species suggested
that the \ion{H}{i} collision rates might be reasonably well described by Drawin's formula with \kH\ $\ge 1$. For example, empirical estimates by \citet{Grattonetal99} resulted in \kH\ = 3 for \ion{O}{i} and \ion{Mg}{i} and \kH\ = 30 for \ion{Fe}{i}.
\citet{h_cool_na_o} and \citet{Pereira2009} inferred \kH\ = 1 from the analysis of the center-to-limb variation of solar \ion{O}{i} triplet $\sim7770$\AA\ lines. The same value of \kH\ = 1 was obtained by \citet{Takeda95} from solar \ion{O}{i} line profile fits. For a review of studies constraining empirically the efficiency of \ion{H}{i} collisions, see \citet{Lambert1993,Holweger1996}, and \citet{mash_lund}.

As a result of applying incomplete model atoms and a different treatment of collisions with hydrogen atoms, no consensus on the expected magnitude of the non-LTE effects was achieved in the previous studies of iron, and results were in conflict with each other in some cases. For example, \citet{Korn03} found a negligible discrepancy between the non-LTE spectroscopic and {\sc Hipparcos} astrometric distances of the halo star HD\,84937, while the non-LTE calculations of \citet{Thevenin1999} resulted in a 34\,\%\ smaller spectroscopic distance of that same star.

This study aims to construct a fairly complete model atom of iron, to be tested using the Sun and selected cool stars with high-quality observed spectra and reliable stellar parameters. Compared with the previous non-LTE analyses of iron, the model atom of \ion{Fe}{i} was extended to high-lying levels predicted by the atomic structure calculations of \citet{Kurucz2009} and this turned out to be crucial for a correct treatment of the SE of iron in cool star atmospheres. With our improved model atom we tried to constrain the scaling factor \kH\ empirically. We realize that the real temperature dependence of hydrogen collision rates could be very different from that of the classical Drawin formalism, and we may not always achieve consistent \kH\ values from the analysis of different stars. We also realize that the required thermalizing process not involving electrons in the atmospheres of cool metal-poor stars could be very different from inelastic collisions with neutral hydrogen atoms. For example, \citet{barklem2010} \citep[see also][]{barklem2003,belyaev03,belyaev2010} uncovered the importance of the ion-pair production and mutual neutralisation process $A(nl) + {\rm H}(1s) \rightleftharpoons A^+ + {\rm H^-}$ for the SE of Li and Na. Since no accurate calculations of either inelastic collisions of iron with neutral hydrogen atoms or other type processes are available, we simulate an additional source of thermalization in the atmospheres of cool stars by parametrized \ion{H}{i} collisions. Investigating the Sun as a reference star for further stellar differential line-by-line analysis, we also derive the solar iron absolute abundance and check the solar \ion{Fe}{i}/\ion{Fe}{ii} ionization equilibrium using an extended list of lines, which can be detected at solar metallicity down to [Fe/H] = $-2.5$. We find it important to inspect the accuracy of atomic data for various subsamples of iron lines in view of comprehensive abundance studies across the Galaxy targeting at stars of very different metallicities.

This paper is organized as follows. The model atom of iron
 and the adopted atomic data are presented in Sect.\,\ref{sect:NLTE}. There we also discuss how including the bulk of predicted \ion{Fe}{i} levels in the model atom affects the SE of iron. In Sect.\,\ref{sect:sun}, the solar iron spectrum is studied to provide the basis for further
differential analyses of stellar spectra. Section\,\ref{sect:stars} describes observations and stellar parameters of our sample of stars, and Sect.\,\ref{sect:stellar_iron} investigates which line-formation assumptions lead to consistent element abundances from both  ions, \ion{Fe}{i} and \ion{Fe}{ii}. Uncertainties in the iron non-LTE abundances are estimated in Sect.\,\ref{sect:uncertainty}. Our recommendations and conclusions are given in Sect.\,\ref{conclusion}.

\section{The method of non-LTE calculations for iron}\label{sect:NLTE}

In this section, we describe the model atom of iron and the programs used for computing the level populations and spectral line profiles.

\begin{figure*}
\hbox{ %\hspace{-6mm}
\resizebox{170mm}{!}{\rotatebox{90}{\includegraphics{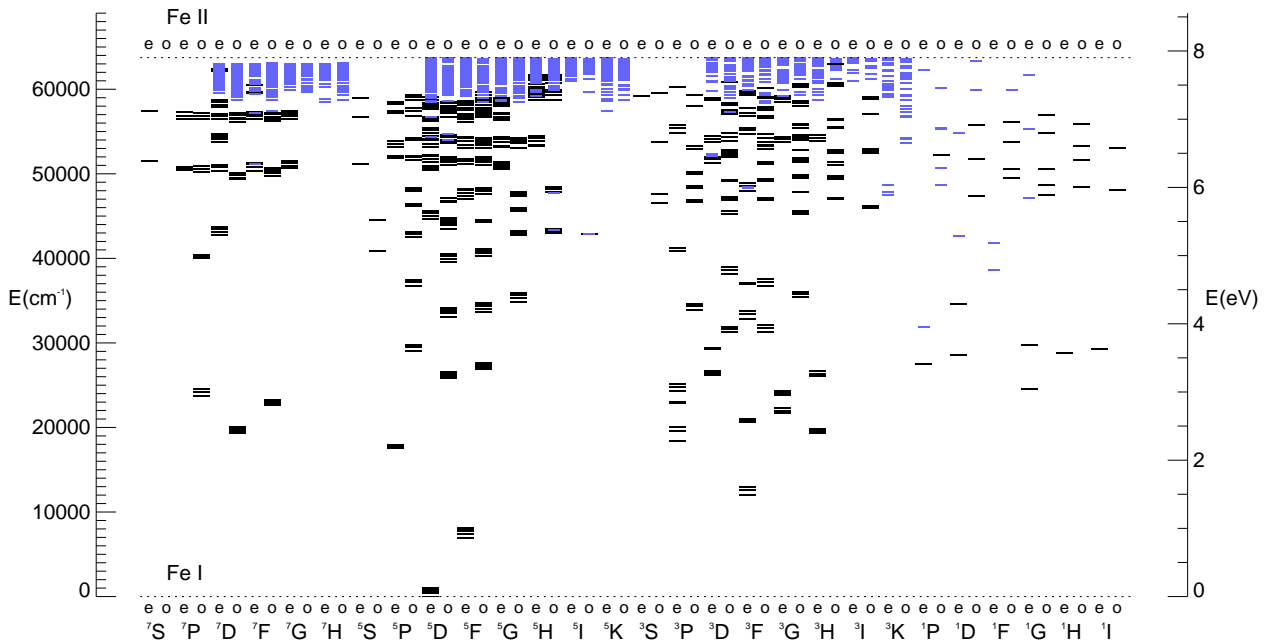}}}}
\vspace{-6mm}
\caption[]{The atomic structure of \ion{Fe}{i} as obtained from laboratory experiments (long lines, black lines in the color online version of the paper) and calculations (short lines, blue lines in the color online version of the paper). See text for the sources of data.}
\label{Fig:fe1_full_atom}
\end{figure*}

\subsection{Model atom}\label{sec:model}

{\it Energy levels.} Iron is almost completely ionized
throughout the atmosphere of stars with an effective temperature above
4500\,K. For example, nowhere in the solar atmosphere, the fraction of \ion{Fe}{i} exceeds 10\%. Such minority species are particularly sensitive to non-LTE
effects because any small change in the ionization rates changes their populations by a large amount. To provide close collisional coupling of \ion{Fe}{i} to the continuum electron reservoir and consequently establish a realistic ionization balance between the neutral
 and singly-ionized species,
%the atomic model for \ion{Fe}{i} has to be fairly complete.
the energy separation of the highest levels in the model atom from the ionization limit must be smaller than the mean kinetic energy of electrons, i.e., 0.5~eV for atmospheres of solar temperature.

In our earlier study \citep[][ hereafter Paper~I]{Gehren2001a}, the model atom of \ion{Fe}{i} was built up using all the energy levels from the experimental analysis of \citet{Nave1994}, in total 846 levels with an excitation energy, \Eexc, up to 7.5~eV. The later updates of \citet{Schoenfeld1995} and the measurements of \citet{Brown1988} provided another 112 energy levels for \ion{Fe}{i}. All the known levels of \ion{Fe}{i} are shown in Fig.\,\ref{Fig:fe1_full_atom} by long horizontal bars. How complete is this system of levels? In the same figure, the upper energy levels predicted by \citet{Kurucz2009} in his calculations of the \ion{Fe}{i} atomic structure are plotted by short bars. As described by \citet{Grupp09}, the new calculations for iron included additional laboratory levels and more configurations than the earlier
work of \citet{Kurucz1992}. It can be seen from Fig.\,\ref{Fig:fe1_full_atom} that the system of measured levels is 
complete below \Eexc\,$\simeq 45\,000$~cm$^{-1}$ (5.6~eV), except perhaps for singlets. However, laboratory experiments do not see most of the high-excitation levels with \Eexc\ $>$ 7.1~eV, which should contribute a lot to provide close collisional coupling of \ion{Fe}{i} levels to the \ion{Fe}{ii} ground state.

\begin{figure*}
\hbox{ \hspace{-3mm}
\resizebox{170mm}{!}{\rotatebox{90}{\includegraphics{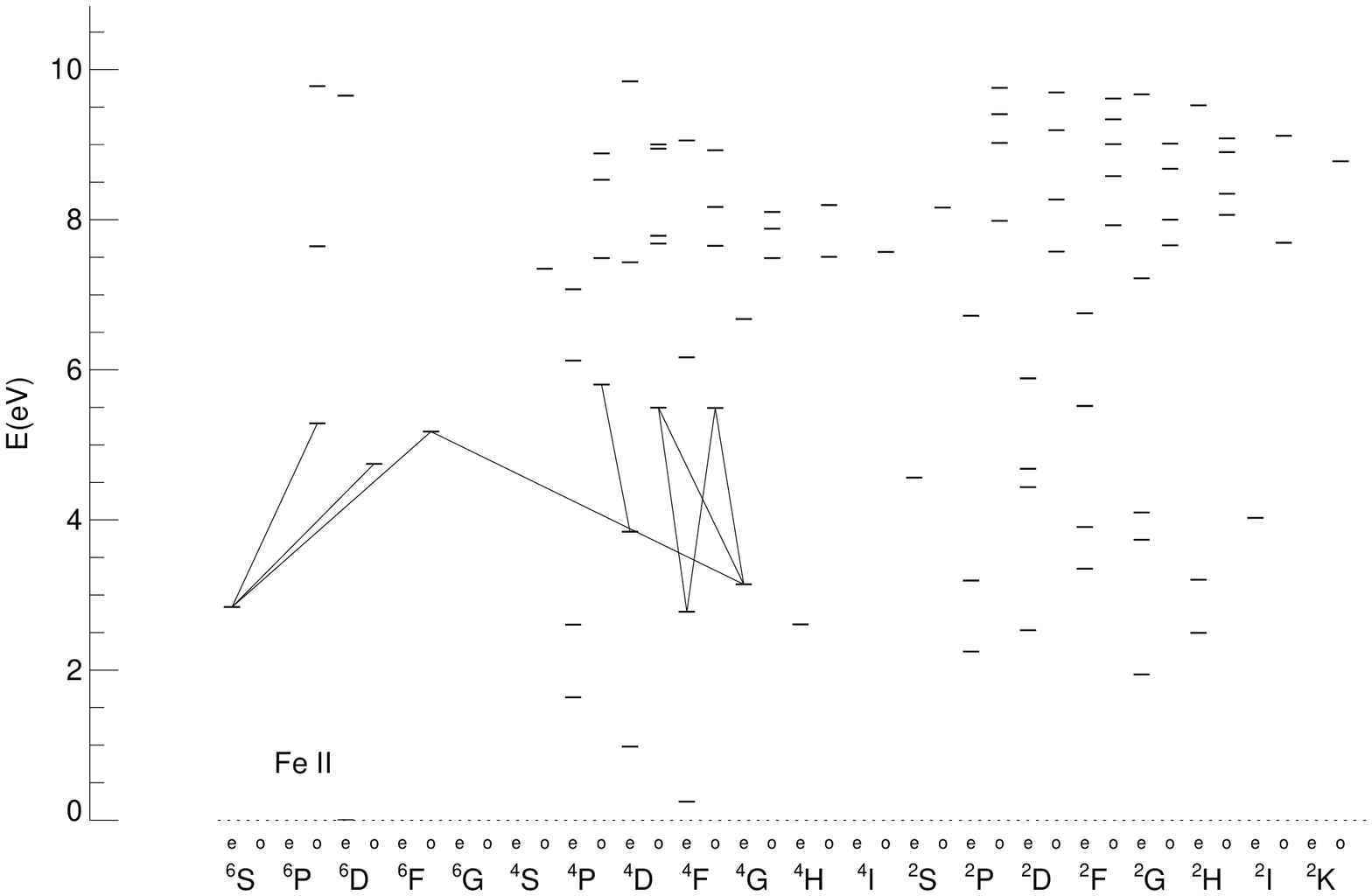}}}}

\vspace{-16mm}
\hbox{ %\hspace{-6mm}
\resizebox{170mm}{!}{\rotatebox{90}{\includegraphics{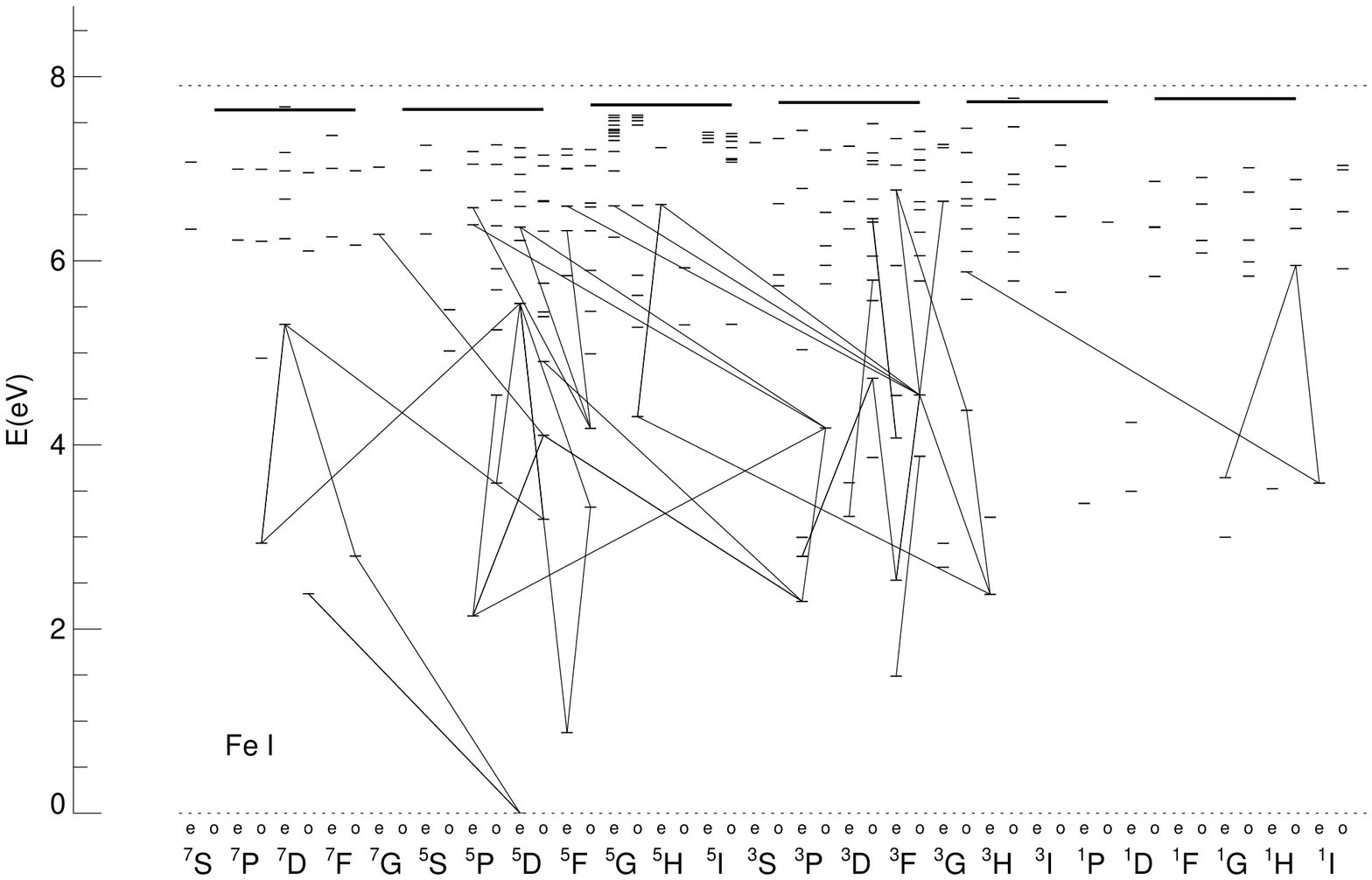}}}}
\vspace{-6mm}
\caption[]{The model atom of \ion{Fe}{i} (bottom panel) and  \ion{Fe}{ii} (top panel). The super-levels are indicated by long lines. The iron spectral lines used in abundance analysis arise from the transitions shown as continuous lines.}
\label{Fig:fe1_atom}
\end{figure*}

In our present study, the model atom of \ion{Fe}{i} is constructed using not only all the known energy levels, but also the
predicted levels with \Eexc\ up to 63\,697~cm$^{-1}$ (7.897~eV), in total 2970 levels. The measured levels belong to 233 terms. Neglecting their multiplet fine structure we obtain 233 levels in the model atom. The predicted and measured levels with common parity and close energies are combined whenever the energy separation is smaller than 150~cm$^{-1}$ at \Eexc\ $<$ 60\,000~cm$^{-1}$ and smaller than 210~cm$^{-1}$ at \Eexc\ $>$ 60\,000~cm$^{-1}$. The remaining predicted levels, all above \Eexc\ = 60\,000~cm$^{-1}$, are used to make up six super-levels. For super-levels, the energy is calculated as a $g-$weighted mean and the total statistical weight amounts from 940 to 2160. Our final model atom of \ion{Fe}{i} is shown in Fig.~\ref{Fig:fe1_atom}.

For \ion{Fe}{ii} (Fig.~\ref{Fig:fe1_atom}), we rely on the reference model atom treated in Paper~I and use the levels belonging to 89 terms with \Eexc\ up to 10~eV. Multiplet fine structure is neglected. The ground state of \ion{Fe}{iii} completes the system of levels in the model atom.

{\it Radiative bound-bound (b-b) transitions.} In total, 11958 allowed transitions occur in our final model atom of \ion{Fe}{i}. Their average-''multiplet'' $gf$-values are calculated using the \citet{Nave1994} compilation for 2649 lines, \citet{Kurucz2009} calculations for the transitions between the measured levels, in total, for 73\,434 lines, and \citet{Kurucz2009} calculations for the transitions between the measured and predicted and between two predicted levels, in total, for 281\,007 lines. The quality of the recent calculations of \citet{Kurucz2009} is estimated by comparing with the \citet{Nave1994} data. The latter can be referred to as experimental data though, for a minority of the lines compiled by \citet{Nave1994}, $gf$-values were derived from solar spectra assuming a solar iron abundance. Taking the experimental sample as reference the theoretical data show a single line scatter of 0.54~dex. Ignoring all lines with deviations above 1~dex (6~\%) the scatter is reduced to 0.33~dex. The reliability of the calculated $gf$-values is thus roughly characterized by a factor of 2 statistical accuracy only. It is, however, important to note that there is no systematic shift between experimental and calculated data. The mean difference is $\Delta\log gf$(calculated - experimental) = $-0.04$ for the whole sample of lines. For 1525 allowed transitions in \ion{Fe}{ii}, $gf$-values are fully based on the data calculated by \citet{Kurucz1992}.

{\it Radiative bound-free (b-f) transitions.} Photoionization is the most important process deciding whether the \ion{Fe}{i} atom tends to depart from LTE in the atmosphere of a cool star. As in Paper~I, our non-LTE calculations rely on the photoionization cross-sections of the IRON project \citep{Bautista1997}. They are available for all the levels of \ion{Fe}{i} in our model atom with the ionization edge in the UV spectral range, in total, for 149 levels. For the remaining levels of \ion{Fe}{i} and all the \ion{Fe}{ii} levels, the hydrogenic approximation was used. We note that photoionization weakly affects the SE of \ion{Fe}{ii} because \ion{Fe}{iii} constitutes only an extremely small fraction of the total iron atoms.

{\it Collisional transitions.} All levels in our model atom are coupled via collisional excitation and ionization by electrons and by neutral hydrogen atoms. For \ion{Fe}{i}, electron-impact excitations are not yet known with sufficient accuracy, and, as in Paper~I, our calculations of collisional
%electron-impact excitation and ionization
rates rely on theoretical approximations. We use the formula of \citet{Reg} for the allowed transitions and assume that the effective collision strength $\Upsilon$ = 1 for the forbidden transitions. The latter assumption can be verified for the single forbidden transition of \ion{Fe}{i}, \eu{a}{5}{D}{}{} - \eu{a}{5}{F}{}{}, for which the calculations of \citet{PB1997} in the IRON Project lead to $\Upsilon$ = 0.98 at $T = 4000$~K.

For the transitions between the \ion{Fe}{ii} terms up to \eu{z}{4}{D}{\circ}{}, we employ the data from the close-coupling calculations of \citet{ZP1995} and \citet{Bautista1996,Bautista1998}. The same formulas as for \ion{Fe}{i} are applied for the remaining transitions in \ion{Fe}{ii}. Ionization by electronic collisions is calculated as in Paper~I from the \citet{Seaton1962} classical path approximation with a mean Gaunt factor set equal to $\overline{g}$ = 0.1 for \ion{Fe}{i} and to 0.2 for \ion{Fe}{ii}.

\begin{figure}
\hbox{ %\hspace{-6mm}
\resizebox{88mm}{!}{\includegraphics{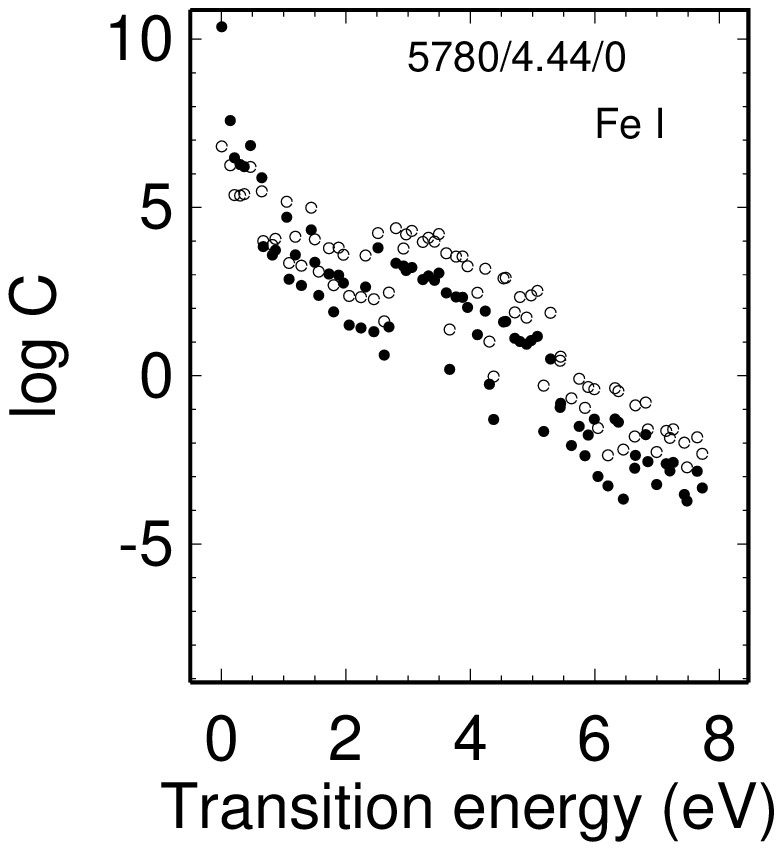}

\hspace{-15mm}
\includegraphics{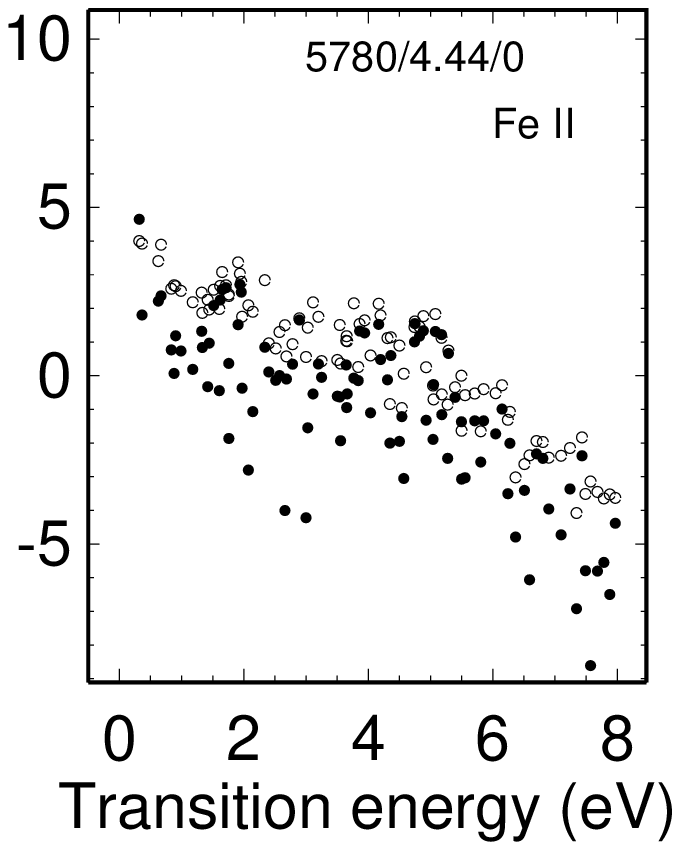}}}
\hbox{ %\hspace{-6mm}
\resizebox{88mm}{!}{\includegraphics{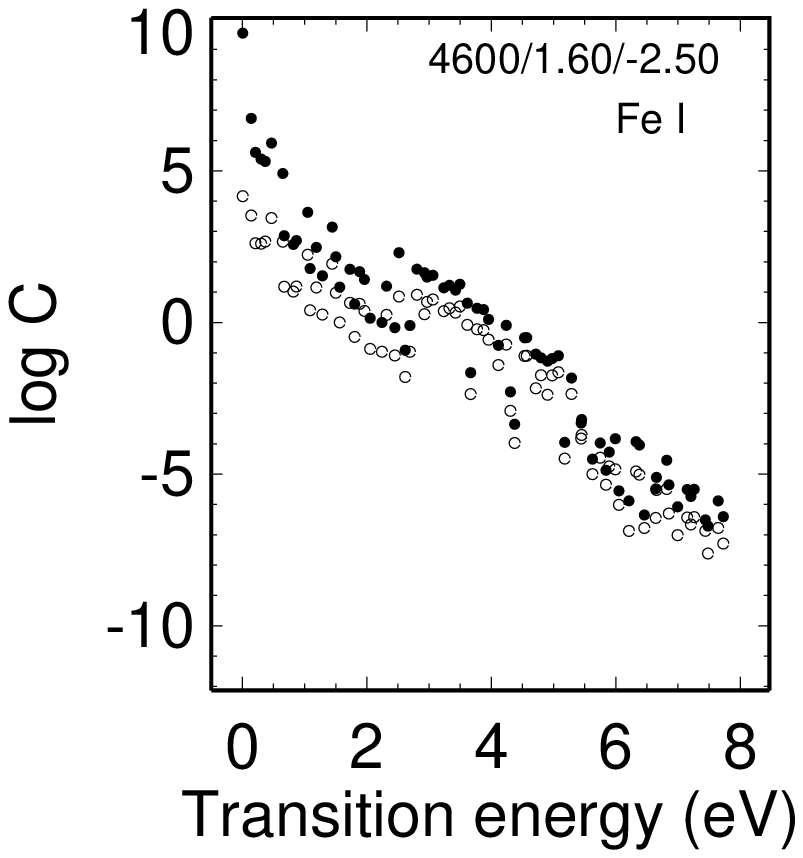}

\hspace{-15mm}
\includegraphics{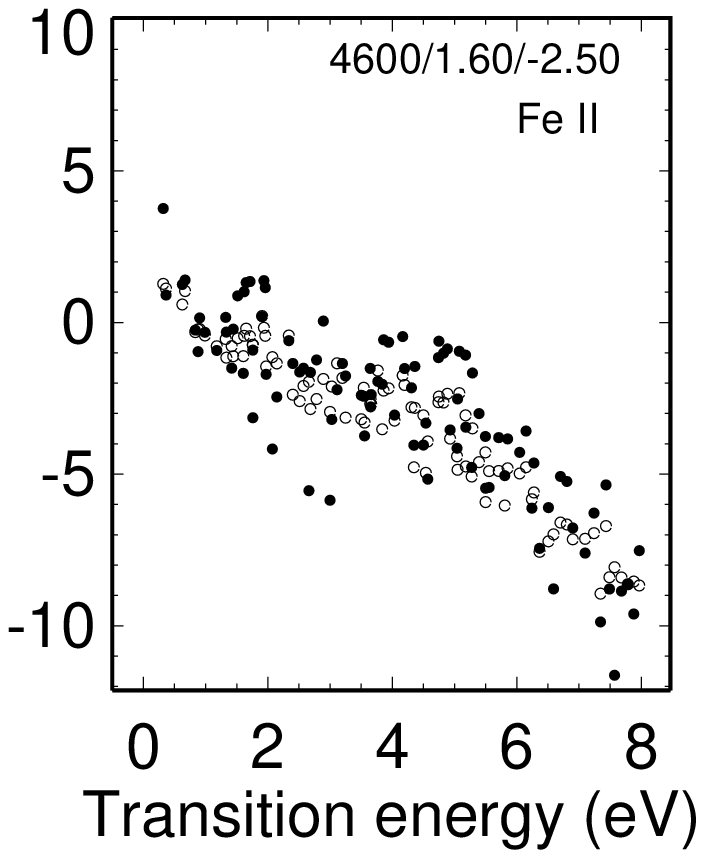}}}
%\vspace{-6mm}
\caption[]{ Electron (open circles) and hydrogen (filled circles) collision rates of representative transitions in \ion{Fe}{i} (left column) and \ion{Fe}{ii} (right column) as a function of the transition energy in the model 5780/4.44/0 at a depth point of $\log\tau_{5000} = -0.5$, where $T$ = 5583~K, $n_H = 0.922\cdot 10^{17}$, and $n_e = 0.137\cdot 10^{14}$ (top row), and in the model $4600/1.60/-2.50$ at a depth point of $\log\tau_{5000} = -0.5$, where $T$ = 4174~K, $n_H = 0.212\cdot10^{17}$, and $n_e = 0.286\cdot 10^{11} $ (bottom row). Everywhere, \kH\ = 0.1 has been assumed.}
\label{Fig:collis}
\end{figure}

For collisions with \ion{H}{i} atoms, we employ the formula of \citet{Steenbock1984} for allowed $b-b$ and $b-f$ transitions and, following \citet{Takeda94}, a simple relation between hydrogen and electron collisional rates, $C_H = C_e \sqrt{(m_e/m_H)} N_H/N_e$, for forbidden transitions. The efficiency of \ion{H}{i} collisions is treated as a free parameter in our attempt to achieve consistent element abundances derived from the two ionization stages, \ion{Fe}{i} and \ion{Fe}{ii}, in the selected stars. For each object, the calculations were performed with a scaling factor of \kH\ = 0, 0.1, 1, and 2. In Fig.\,\ref{Fig:collis}, we compare the electron-impact excitation rates with the corresponding \ion{H}{i} collision rates for representative transitions in \ion{Fe}{i} and \ion{Fe}{ii} in the line-forming layers ($\log \tau_{5000} = -0.5$) of the solar-metallicity ($\Teff$/$\logg$/[Fe/H]\footnote{In the classical notation, where [X/H] = $\log(N_{\rm X}/N_{\rm H})_{star} - \log(N_{\rm X}/N_{\rm H})_{Sun}$.} = 5780/4.44/0) and metal-poor (4600/1.60/$-2.50$) models. At the selected depth point, the electron number density $n_e$ in the solar-metallicity model is by a factor of 500 higher than that in the metal-poor and cool atmosphere, while the difference in neutral hydrogen number density $n_H$ is much smaller, only a factor of 4. As a consequence, the hydrogen collisions dominate the total collisional rate for most \ion{Fe}{i} and \ion{Fe}{ii} transitions in the metal-poor model. For solar-metallicity models, the influence of inelastic \ion{H}{i} collisions on the SE of iron is expected to be weaker because $C_H/C_e > 1$ is fulfilled only for the \ion{Fe}{i} transitions with small energy separation, i.e., smaller than 0.5~eV. For \ion{Fe}{ii}, the close-coupling electron-impact excitation rates of \citet{ZP1995} and \citet{Bautista1996,Bautista1998} are higher compared to the hydrogen collision rates even for small transition energies. A similar relation between $C_H$ and $C_e$ also holds outside $\log \tau_{5000} = -0.5$ in the solar-metallicity model. In the metal-poor model, the weakest iron lines form inside $\log \tau_{5000} = -0.5$, where the role of hydrogen collisions is weakened due to a decreasing $n_H/n_e$ ratio.

\subsection{Programs and model atmospheres}

\begin{figure*}
%\vspace*{-3mm}
%\hbox{
%\hspace{-6mm}
%\resizebox{88mm}{!}{\includegraphics{sun_flux_lte_1500to3000.ps}}
\resizebox{170mm}{!}{\includegraphics{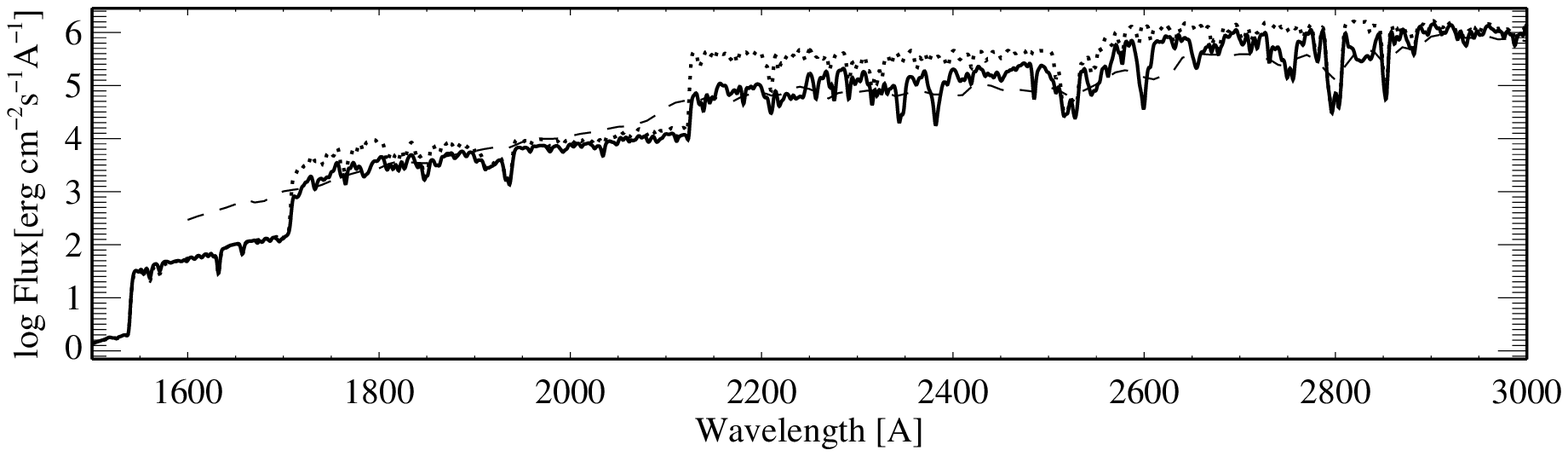}}

\vspace{-5mm}
\resizebox{170mm}{!}{\includegraphics{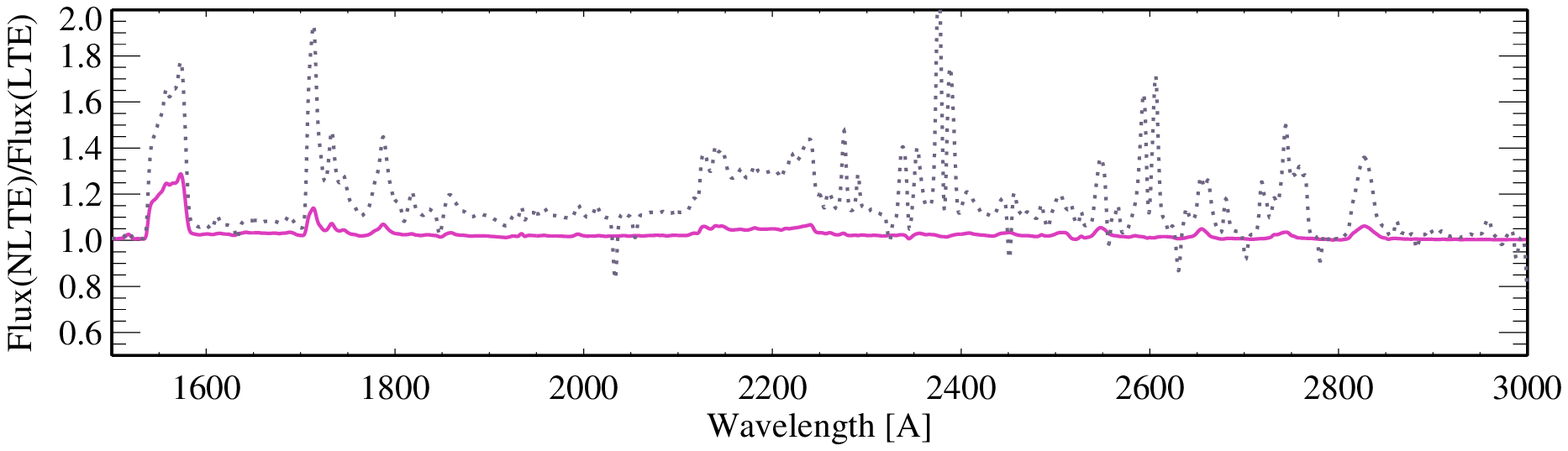}}
%\vspace{5mm}
\caption[]{Top panel: emergent fluxes calculated using the model of $\Teff$ = 5777~K, $\logg$ = 4.44, and [M/H] = 0 with iron ($\eps{Fe} = 7.50$, non-LTE with \kH\ = 0.1, continuous curve) and without iron (dotted curve) compared to the solar flux taken from \citet[][, dashed curve]{Woods1996}. The observed flux below 1700~\AA\ is contributed by the solar chromosphere. Bottom panel: the ratio between the non-LTE (\kH\ = 0.1) and LTE emergent fluxes in the same model atmosphere with $\eps{Fe} = 7.50$ from the calculations with the final (continuous curve) and reduced (dotted curve, see Sect.\,\ref{sect:atom_compl}) model atoms. For clearer illustration, the theoretical fluxes were convolved with a Gaussian profile of $\Delta\lambda_D = 2$~m\AA.} \label{Fig:sun_flux}
\end{figure*}

We compute statistical equilibrium populations for \ion{Fe}{i} and  \ion{Fe}{ii} while keeping the atmospheric structure fixed. This is justified by the following considerations. Despite the fact that iron is an important source of the UV continuous opacity in an atmosphere of close-to-solar metallicity (see Fig.\,\ref{Fig:sun_flux}, top panel),
%where the emergent fluxes in the model of $\Teff$ = 5777~K, $\logg$ = 4.44, and [M/H]\footnote{In the classical notation, where [X/H] = $\log(N_{\rm X}/N_{\rm H})_{star} - \log(N_{\rm X}/N_{\rm H})_{Sun}$.} = 0 with and without iron in the atmosphere are shown. At the same time,
 the variations in its excitation and ionization state between LTE and non-LTE are found to have, at the stellar parameters with which we are concerned, no significant effect on the emergent fluxes (Fig.\,\ref{Fig:sun_flux}, bottom panel). Therefore, only minor effects on atmospheric temperature and electron number density distribution are expected.

 The solution of the non-LTE problem with such a comprehensive model atom as treated in this study is only possible, at present, with classical plane-parallel (one-dimensional, 1D) model atmospheres.
All calculations were performed with
%plane-parallel and blanketed LTE
model atmospheres computed with the MAFAGS-OS code \citep{Grupp04,Grupp09}, which is based on up-to-date continuous opacities and
includes the effects of line-blanketing by means of opacity sampling.

We used a revised version of the DETAIL program \citep{detail} based on the accelerated lambda iteration (ALI) scheme
following the efficient method described by \citet{rh91,rh92} in order to solve the coupled
radiative transfer and statistical equilibrium equations. The opacity package of the DETAIL code has been updated by the inclusion of the quasi-molecular Lyman $\alpha$
satellites following the implementation by \citet{cast_kur2001}
of the \citet{allard98} theory and by the use of the Opacity
Project \citep[see][for a general review]{OP} photoionization cross-sections for the calculations of $b-f$ absorption of \ion{C}{i}, \ion{N}{i}, \ion{Mg}{i}, \ion{Si}{i}, \ion{Al}{i}, and \ion{Fe}{i}. In addition to the continuous background opacity, the line opacity
introduced by \ion{H}{i} and metal lines was taken into
account by explicitly including it in solving the radiation
transfer. The metal line list was extracted from the
\citet{cdrom18} compilation and VALD database \citep{vald}. It includes about 720\,000 atomic
and molecular lines between 500 and 300\,000\,\AA. The \ion{Fe}{i} and \ion{Fe}{ii} lines as well as $b-f$ absorption processes were excluded from the background.

All the $b-b$ and $b-f$ transitions of
\ion{Fe}{i} and \ion{Fe}{ii} were explicitly taken into account in
the SE calculations. The 57 strongest $b-b$ transitions
were treated using Voigt profiles and the remaining transitions using depth-dependent Doppler profiles. Microturbulence was accounted for
by inclusion of an additional term in the Doppler width.

The departure coefficients, $b_i = n_i^{\rm NLTE}/n_i^{\rm LTE}$, were then used to compute synthetic line profiles via the {\sc SIU} program \citep{Reetz}. Here,
$n_i^{\rm NLTE}$ and $n_i^{\rm LTE}$ are the statistical
equilibrium and thermal (Saha-Boltzmann) number densities,
respectively. In this step of
the calculations, Voigt profile functions were adopted and the same
microturbulence value \Vmic\ as in DETAIL was applied.

\subsection{Effect of model atom completeness on the statistical equilibrium of iron}\label{sect:atom_compl}

\begin{figure*}
%\vspace*{-3mm}
\hbox{
%\hspace{-6mm}
\resizebox{170mm}{!}{\rotatebox{0}{\includegraphics{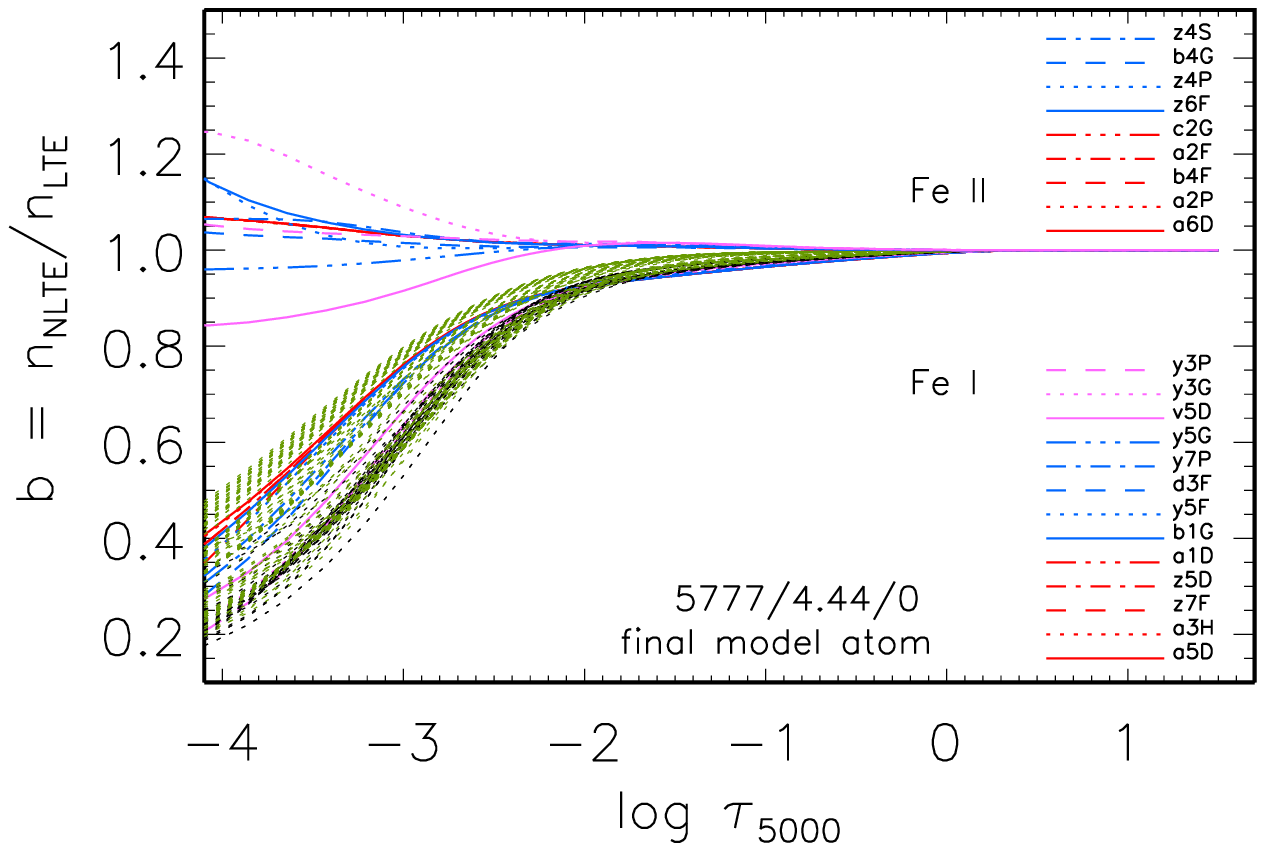}}
\hspace{-8mm} \rotatebox{0}{\includegraphics{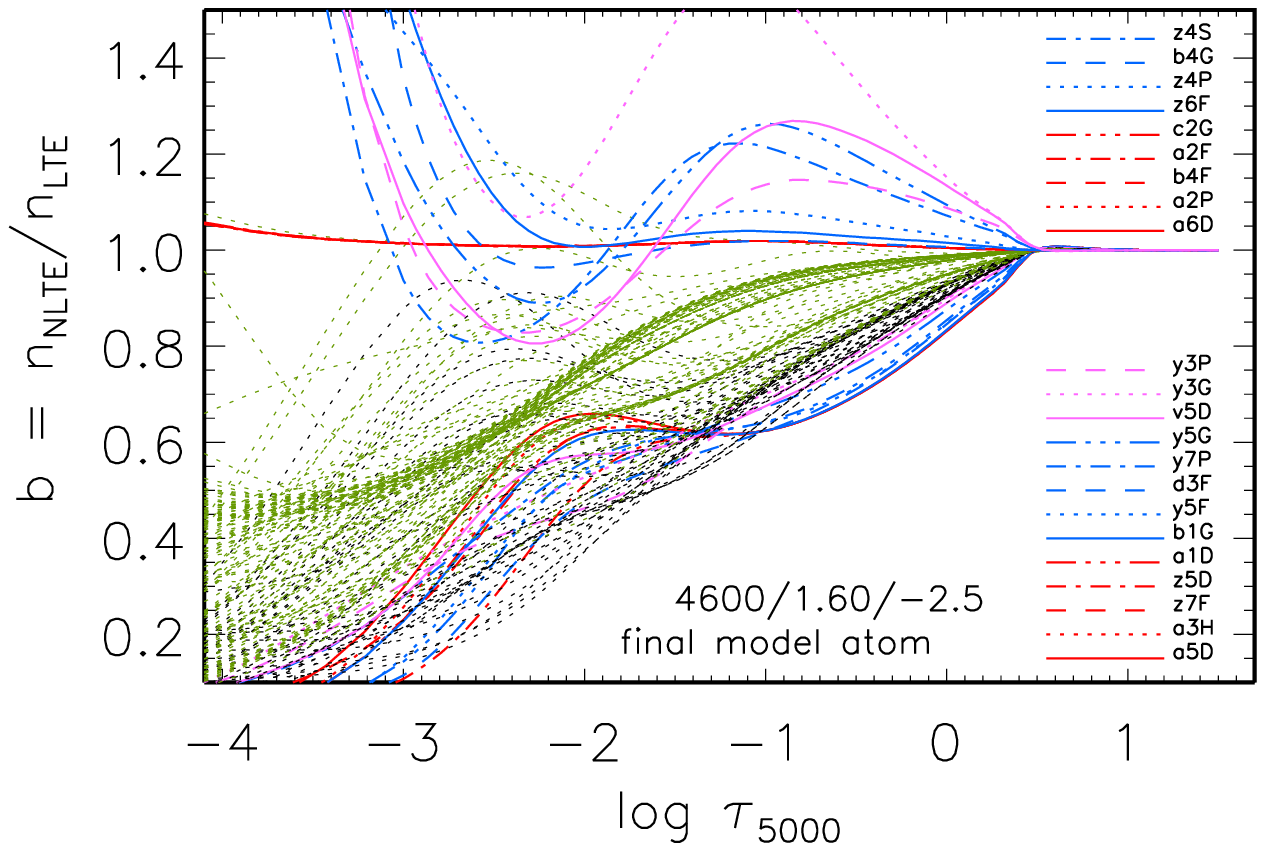}}}}
\vspace{-5mm}
\hbox{
%\hspace{-6mm}
\vspace*{-3mm}
\resizebox{170mm}{!}{\rotatebox{0}{\includegraphics{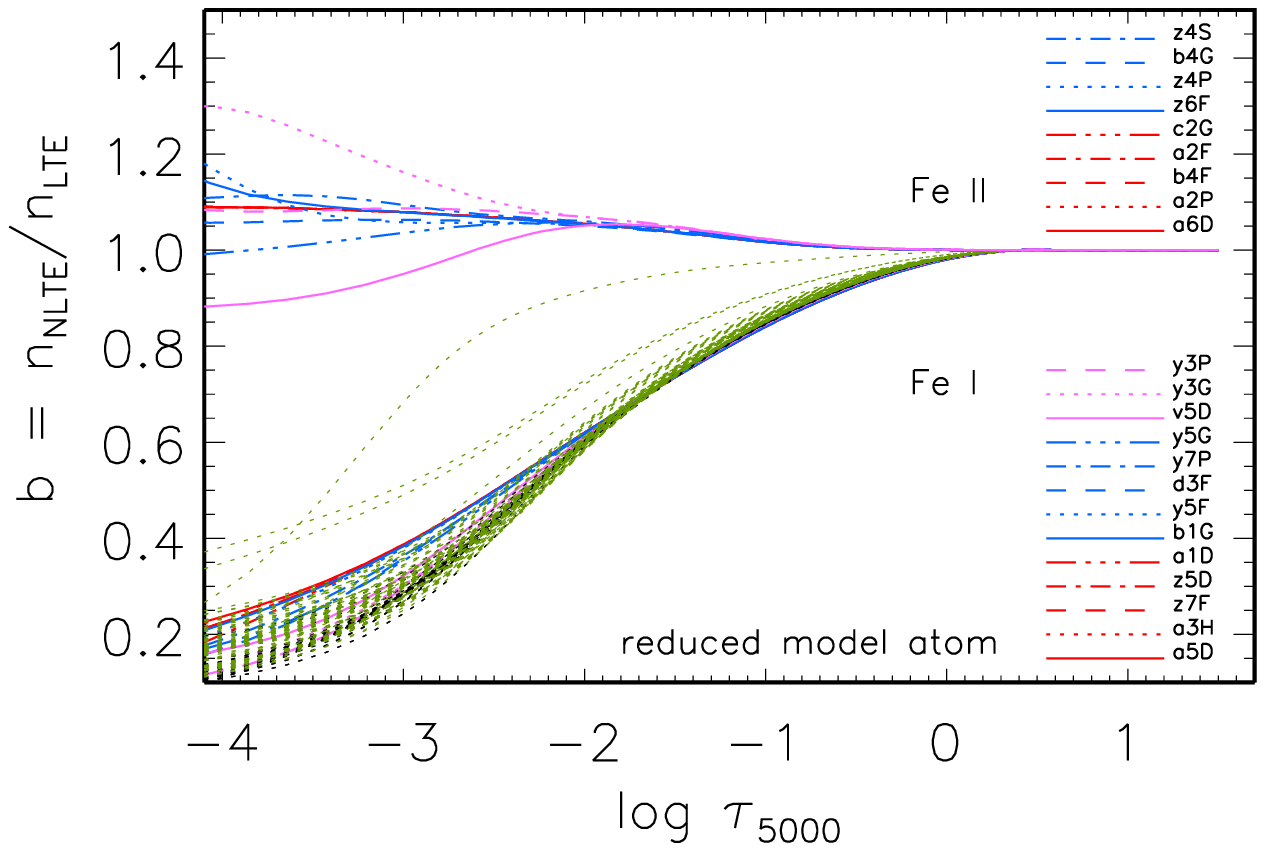}}
\hspace{-8mm} \rotatebox{0}{\includegraphics{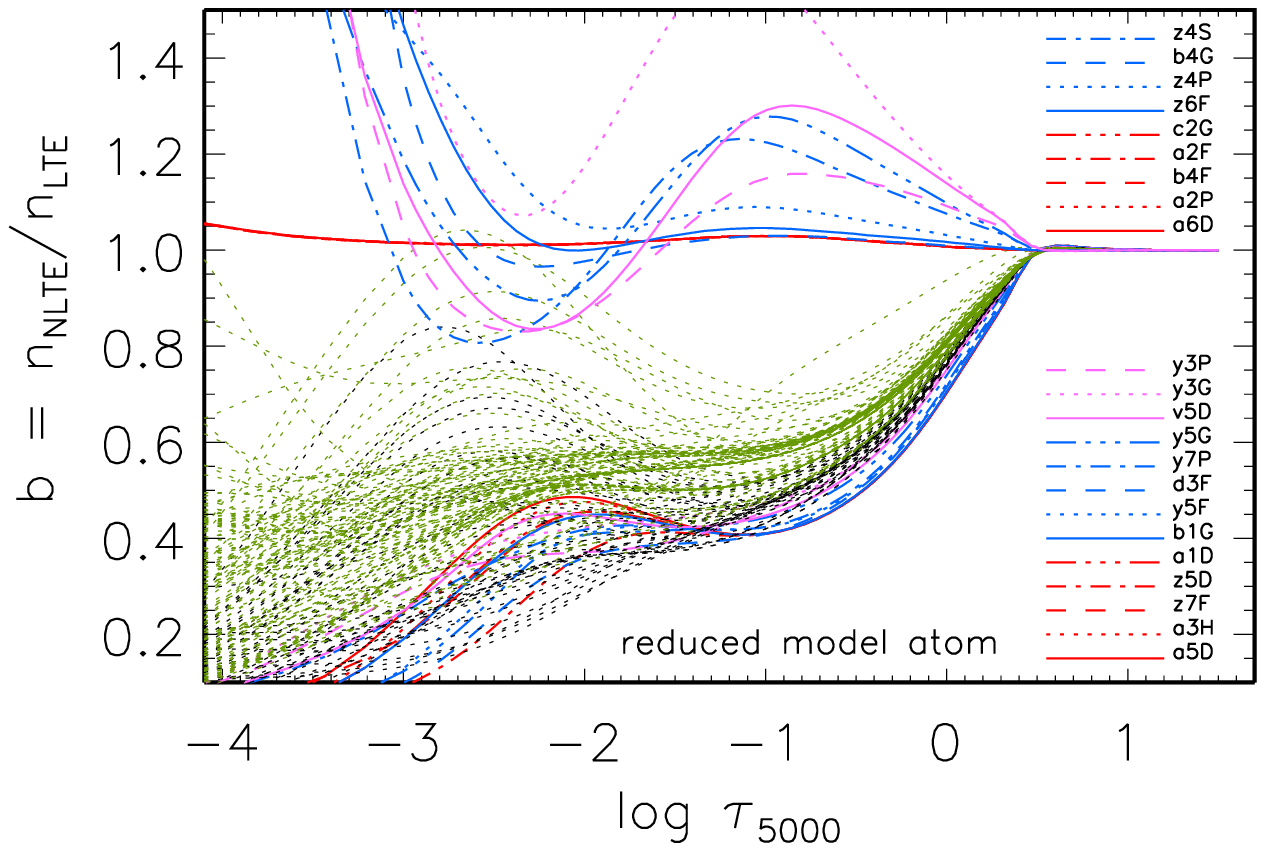}}}}
%\vspace{5mm}
\caption[]{Departure coefficients, $b$, for the levels of \ion{Fe}{i} and \ion{Fe}{ii} as a function of $\log \tau_{5000}$ in the model atmospheres 5777/4.44/0 (left column) and 4600/1.60/$-2.50$ (right column) from the calculations with our final model atom of iron (top row) and the reduced model atom which ignores the predicted levels of \ion{Fe}{i}. Every fifth of the first 60 levels of \ion{Fe}{i} is shown. They are quoted in the bottom right part of each panel. All the remaining higher levels of \ion{Fe}{i} are plotted by dotted curves. For \ion{Fe}{ii}, every fifth of the first 60 levels is shown. They are quoted in the top right part of each panel. The letters z, y, x, ... are used to denote the odd parity terms and a, b, c, ... for the even parity terms.
To distinguish \ion{Fe}{i} and \ion{Fe}{ii} levels, note that $b < 1$ for each \ion{Fe}{i} level outside $\log \tau_{5000} = 0$ in the solar metallicity model and $\log \tau_{5000} = 0.4$ in the metal-poor one. An exception is the highest levels in the layers with $\log \tau_{5000}$ around $-2.5$ in the model 4600/1.60/$-2.50$. For \ion{Fe}{ii}, $b \ge 1$ inside $\log \tau_{5000} = -2$ in both model atmospheres.
Everywhere, \kH\ = 0.1. } \label{Fig:bf}
\end{figure*}

The departure coefficients for selected levels of \ion{Fe}{i} and \ion{Fe}{ii} in two model atmospheres 5777/4.44/0 and 4600/1.60/$-2.50$ are presented in Fig.\,\ref{Fig:bf} (top row).
Our calculations support the results of the previous non-LTE studies of iron \citep{Athay1972,Boyarchuketal85,Takeda91,Grattonetal99,Thevenin1999,Gehren2001a,ShTB01,sh_Fe_2005,Colletal05} in the common stellar parameter range
%, $\Teff = 4600 - 6500$\,K, $\logg = 1.6 - 4.5$, [Fe/H] from $-2.5$ to $+0.1$,
in the following aspects.
\begin{itemize}
\item The non-LTE mechanisms for \ion{Fe}{i} and \ion{Fe}{ii} are 
%general behavior of the departure coefficients of the \ion{Fe}{i} and \ion{Fe}{ii} levels is
independent of effective temperature, surface gravity, and metallicity.
\item All the levels of \ion{Fe}{i} are underpopulated in the atmospheric layers above $\log \tau_{5000} = 0$. The main non-LTE mechanism is the overionization caused by superthermal radiation of a non-local origin below the thresholds of the levels with excitation energy of 1.4 to 4.5~eV.
\item \ion{Fe}{ii} dominates the element number
density over all atmospheric depths. Thus, no process seems to affect the \ion{Fe}{ii} ground-state and low-excitation-level populations significantly, and they keep their thermodynamic equilibrium values.
\end{itemize}

In this study, progress was made in establishing close collisional coupling of
the \ion{Fe}{i} levels near the continuum to the ground state of \ion{Fe}{ii}, due to including in the model atom a bulk of the predicted high-excitation levels of \ion{Fe}{i}. There is no need anymore in the enforced upper level thermalization procedure that was applied by \citet{Gehren2001a,Korn03}, or \citet{Colletal05}.

To illustrate the difference between the present and previous non-LTE studies, we performed test calculations with the atomic model, where the predicted levels of \ion{Fe}{i} are removed. Hereafter, the latter model atom is referred to as the reduced model. It includes 233 terms and 10\,740 radiative $b-b$ transitios in \ion{Fe}{i} and exactly the same levels of \ion{Fe}{ii} as in our final model atom. The bottom row panels of Fig.\,\ref{Fig:bf} show the departure coefficients computed with the reduced model atom. It is evident that
%s representing the atmospheres of the Sun and cool metal-poor giant
(i) the \ion{Fe}{i} terms are more underpopulated in the line-formation layers compared to the populations obtained with our final model atom  (Fig.\,\ref{Fig:bf}, top row), (ii) the highest levels do not couple thermally to the ground state of \ion{Fe}{ii}, and (iii) the majority of the \ion{Fe}{i} terms are close together. The explanation is that the net ionization-recombination rate of the \ion{Fe}{i} levels separated by no more than 0.4~eV from the continuum is much lower in the reduced model atom than in the final model atom, for example, by a factor of 1000 and 150 at $\log\tau_{5000} \simeq -0.6$ in the model atmospheres 5777/4.44/0 and 4600/1.60/$-2.50$, respectively.

 We found that emergent fluxes calculated with the MAFAGS-OS solar model atmosphere and using our final model atom reproduce well the observed solar fluxes of \citet{Woods1996} in the 1700 - 3000\,\AA\ spectral range, where neutral iron is an important source of the continuous opacity (Fig.\,\ref{Fig:sun_flux}, top panel). The observed flux below 1700~\AA\ is contributed by the solar chromosphere. As demonstrated in the bottom panel of Fig.\,\ref{Fig:sun_flux}, the difference in emergent fluxes between LTE and non-LTE is negligible when employing the final model with \kH\ = 0.1, while the use of the reduced model atom leads to a weaker $b-f$ opacity of neutral iron compared to the LTE value, such that, in the 1600 - 2900\,\AA\ spectral range, the change in emergent fluxes is about 20\,\%.

%For the solar-metallicity model atmosphere, the use of {\bf our final model atom results in compared to the LTE values. This is  plotted for As shown by \citet{Prieto2003} and \citet{Grupp09}, the emergent UV fluxes from the solar theoretical 1D+LTE models are consistent with the observed solar fluxes,
%in the spectral ranges overlapping with that plotted in Fig.\,\ref{Fig:sun_flux},
%while this is not the case for the fluxes in the non-LTE model of \citet[][, their Fig.~5]{Prieto2003}.
%Why was the non-LTE model worse than the LTE one?
%A reason was, most probably, in employing an incomplete model atom of \ion{Fe}{i} in the \citet{Prieto2003} study. As shown in our calculations,  }
%\tcol{A} similar difference between non-LTE and LTE emergent UV fluxes\tcol{,} found by \citet[][their Fig.~5]{Prieto2003} was }

In Sect.\,\ref{sect:uncertainty}, we evaluate the difference in solar and stellar non-LTE abundances when using the final and reduced model atoms.

\section{Analysis of the solar \ion{Fe}{i} and \ion{Fe}{ii} lines}\label{sect:sun}

\begin{figure}
%\vspace*{-3mm}
%\hbox{
%\hspace{-6mm}
\resizebox{88mm}{!}{\includegraphics{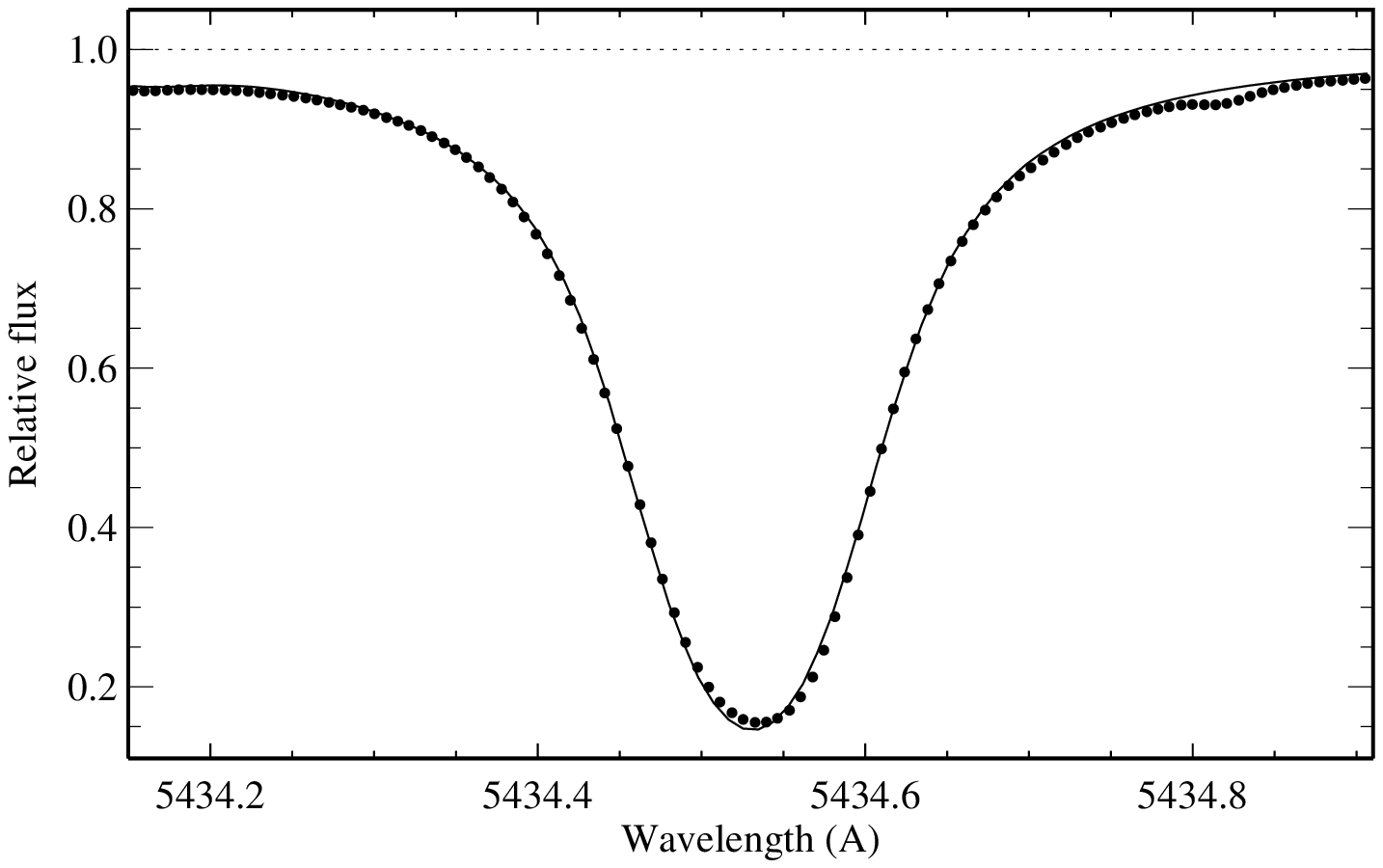}}

\vspace{-5mm}
\resizebox{88mm}{!}{\includegraphics{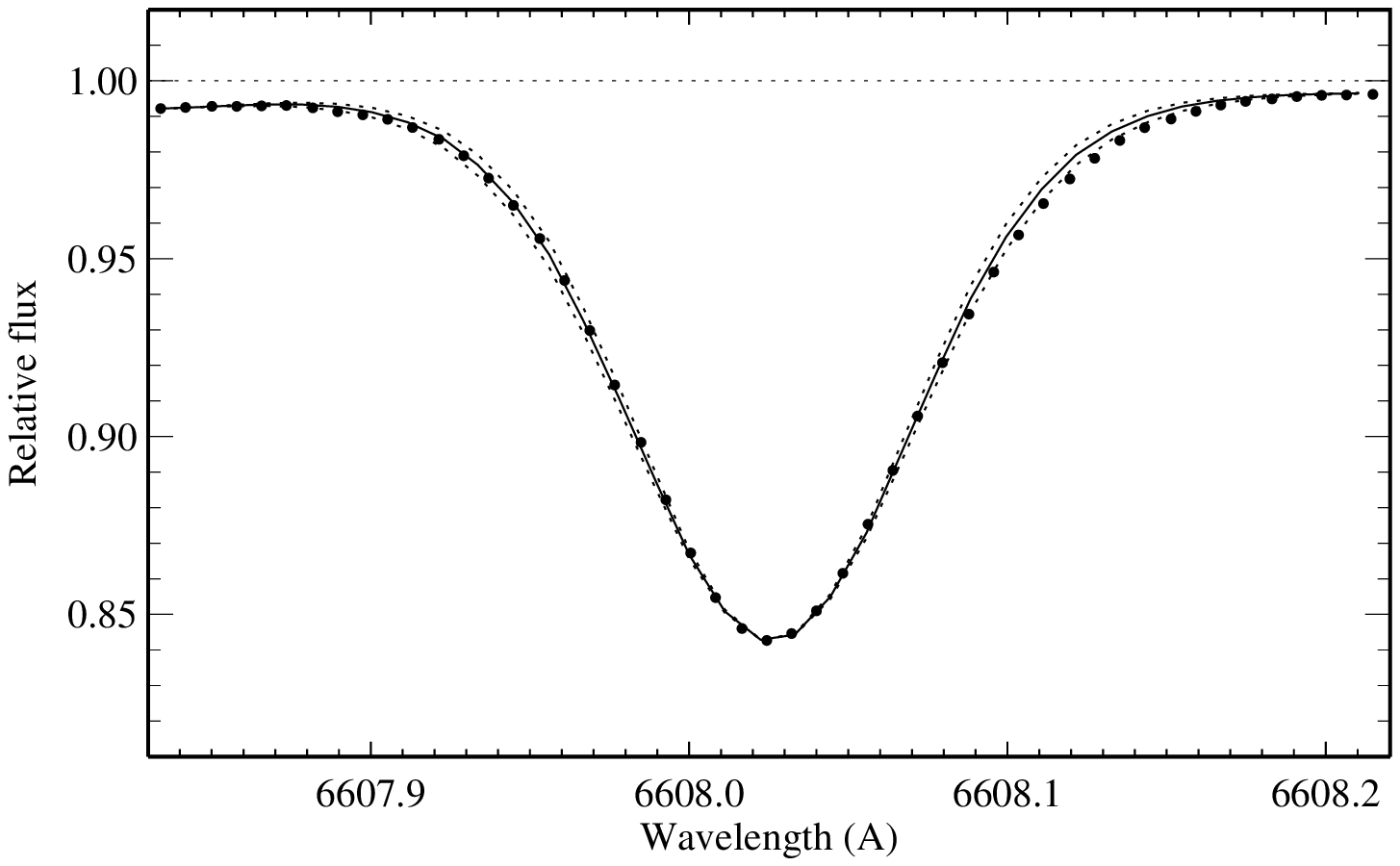}}
%\vspace{5mm}
\caption[]{ The best fits (continuous curve) of the solar \ion{Fe}{i} 5434 and 6608~\AA\ lines (bold dots in top and bottom panels, respectively) obtained from the non-LTE calculations with \kH\ = 0.1. Fitting parameters are given in Table\,\ref{linelist} (online material). For \ion{Fe}{i} 6608~\AA, dotted curves show the effect on the theoretical profile of a $\pm$0.02~dex variation in the final Fe abundance. The macroturbulence velocity was reduced by 0.3\,\kms\ for $\Delta\eps{} = -0.02$~dex and increased by 0.2\,\kms\ for $\Delta\eps{} = +0.02$~dex. The bottom panel illustrates that errors due to an abundance variation in fitting the solar iron line profiles are small.} \label{Fig:line_solar}
\end{figure}

In this section, we derive the solar Fe abundance from the \ion{Fe}{i} and \ion{Fe}{ii} lines. The Sun is used as a reference star for further stellar abundance analysis. The solar flux observations are taken from the Kitt Peak Solar Atlas \citep{Atlas}. The calculations were performed with the MAFAGS-OS model atmosphere 5777/4.44/0 \citep{Grupp09}. A depth-independent microturbulence of 0.9\,\kms\, was adopted. Everywhere in this study, the element abundance is determined from line profile fitting. The uncertainty in the fitting procedure was estimated to be less than 0.02~dex for weak and moderate strength lines (see Fig.\,\ref{Fig:line_solar}) and less than 0.03~dex for strong lines.
Our synthetic flux profiles are convolved with a profile that combines
a rotational broadening of 1.8\,\kms\ and broadening by
macroturbulence with a radial-tangential profile. The most probable macroturbulence velocity $\Vmac$ varied mainly between 2.6 and 3.3\,\kms\ for different lines of neutral iron and between 3.4 and 3.8\,\kms\ for \ion{Fe}{ii} lines.
For comparison, \citet{Gray1997} found solar macroturbulence velocities varying between 2.9 and 3.8\,\kms\ for a small sample of the solar \ion{Fe}{i} lines. The $\Vmac$ values obtained from the non-LTE (\kH\ = 0.1) fits of individual solar iron lines are indicated in Table\,\ref{linelist} (online material).

\subsection{Line selection and atomic data}

 The investigated lines were selected according to the following criteria.
\begin{itemize}
\item The lines should be almost free of visible/known blends in the Sun.
\item For each star, the spectrum should include lines of various strength to provide an abundance analysis of both close-to-solar metallicity and very metal-poor stars.
\item The list of the \ion{Fe}{i} lines should cover as large as possible a range of excitation energies of the lower level to investigate the excitation equilibrium of neutral iron.
\end{itemize}

\noindent
 The selected lines are listed in Table\,\ref{linelist} (online material), along with the transition information and references to the adopted $gf-$values. Van der Waals broadening of the iron lines is accounted for using the most accurate data available from calculations of \citet{omara_sp,omara_pd,omara_df}, and \citet{omara-fe2}.
%The exception is \ion{Fe}{i} 5491\,\AA\ with $\Vmac$ = 3.8\,\kms.
% In fitting at the LTE assumption or non-LTE with \kH\ $> 0.1$, $\Vmac$ changed by no more than 0.1\,\kms. A change by up to 0.2\,\kms\ in $\Vmac$ was required to obtain the best fit in non-LTE with \kH\ = 0.

Despite the existence of many sources of $gf-$values for neutral iron, there is no single source that provides data for all the selected \ion{Fe}{i} lines. We employ experimental $gf-$values from \citet{fe-BKK91,fe-BK94,BIP79,BPS82a,BPS82b,FMW88}, and \citet{fe-OWL91}.
The accuracy of available sets of $gf-$values for \ion{Fe}{i} was discussed by \citet{GS1999}, and the influence of using various sets of data on the derived solar iron abundance was inspected by \citet{Gehren2001b}. Both papers recommended to employ the $gf-$values published by the Hannover \citep{fe-BKK91,fe-BK94} and Oxford \citep{BIP79,BPS82a,BPS82b} groups. Unfortunately, these groups did not provide data for some \ion{Fe}{i} lines important for the stellar iron spectrum analysis.
%It is worth noting that the mean \ion{Fe}{i} based abundance obtained with these data sources is consistent with that based on the {\sc VALD} database $gf-$values \citep[][Table~\ref{Tab:SunAbund}]{vald}.
Five sets of oscillator strengths from \citet{fe-MB09,Moity83,RU98,SSK04}, and the {\sc VALD} database \citep{vald} were inspected for \ion{Fe}{ii}.
We found that the solar mean abundance derived from the \ion{Fe}{ii} lines varies between $\eps{} = 7.41\pm0.11$ and $\eps{} = 7.56\pm0.05$ depending on the  source of $gf-$values (Table~\ref{Tab:SunAbund}). Hereafter, the statistical error is the dispersion in the single line measurements about the mean: $\sigma = \sqrt{\Sigma(\overline{x}-x_i)^2/(n-1)}$.
As reported by \citet{GS1999}, the data of \citet{RU98} for the visible \ion{Fe}{ii} lines were underestimated by 0.11~dex, on average, compared to the $gf-$values obtained using lifetime and branching fraction measurements.

\begin{table} %[htbp]
 \centering
 \caption{\label{Tab:SunAbund} Solar mean iron non-LTE (\kH\ = 0.1) abundances determined using various sources of $gf-$values. }
  \begin{tabular}{lrccl}
   \hline \noalign{\smallskip}
Species       & N$^*$ & $\eps{}$ & $\sigma$  & Source \\
\noalign{\smallskip} \hline \noalign{\smallskip}
 \ion{Fe}{i}  & 54 & 7.56 & 0.09 & Table \ref{linelist} \\
 \ion{Fe}{i}  & 54 & 7.56 & 0.09 & {\sc VALD}		\\
 \ion{Fe}{ii} & 18 & 7.56 & 0.05 & Table \ref{linelist} \\
 \ion{Fe}{ii} & 18 & 7.45 & 0.07 & {\sc VALD}		\\
 \ion{Fe}{ii} & 18 & 7.47 & 0.05 & MB09 		\\
 \ion{Fe}{ii} & 16 & 7.56 & 0.06 & M83  		\\
 \ion{Fe}{ii} & 8  & 7.41 & 0.11 & SSK04		\\
\hline
\multicolumn{5}{l}{ \ $^*$ the number of lines.} \\
%\multicolumn{5}{l}{ \ $^{**}$ the dispersion in the single line measurements} \\
% & \multicolumn{4}{l}{about the mean.} \\
\multicolumn{5}{l}{Sources: {\sc VALD} = \citet{vald},} \\
\multicolumn{5}{l}{MB09 = \citet{fe-MB09},} \\
\multicolumn{5}{l}{M83 = \citet{Moity83},} \\
\multicolumn{5}{l}{SSK04 = \citet{SSK04}.} \\
  \end{tabular}
\end{table}

\subsection{Abundance analysis}

\begin{figure*}
%\vspace*{-3mm}
\hbox{ %\hspace{-6mm}
\resizebox{170mm}{!}{\rotatebox{0}{\includegraphics{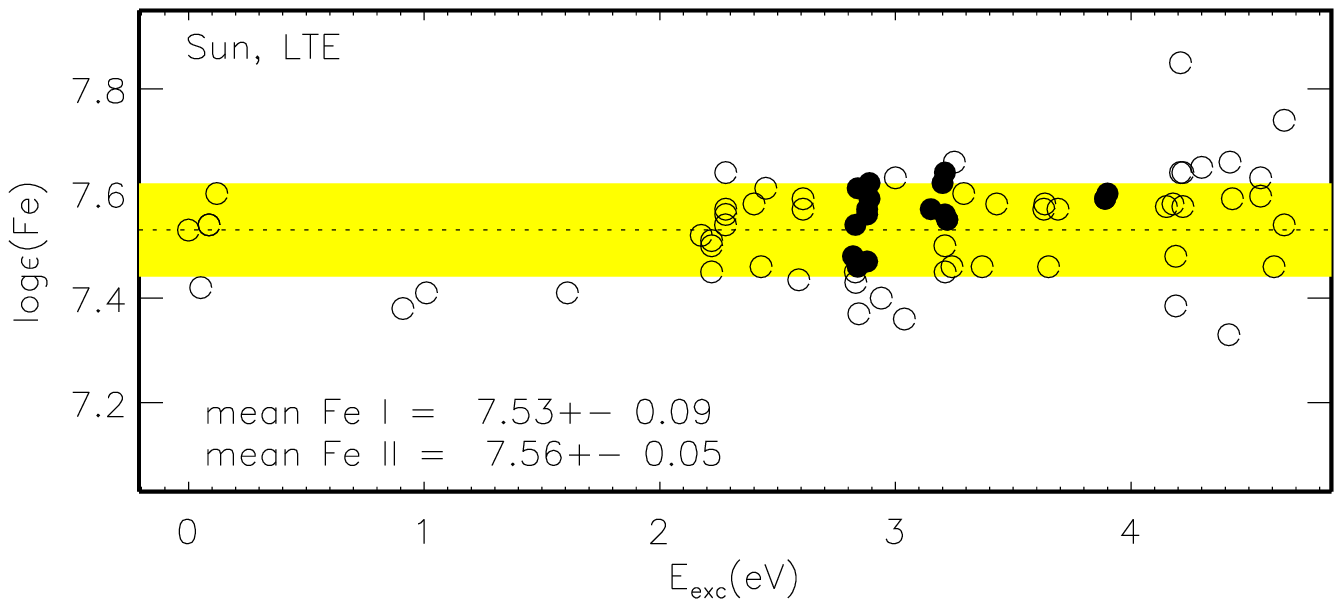}}
\hspace{-8mm}
\rotatebox{0}{\includegraphics{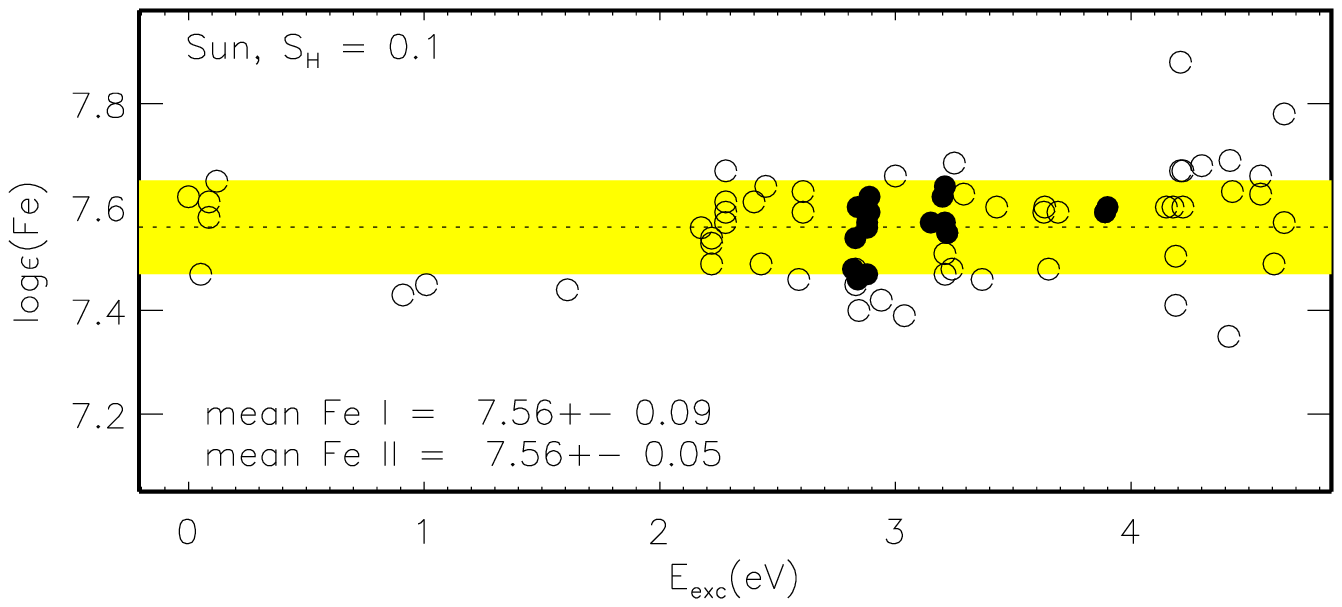}}}}
\vspace{-5mm}
\hbox{ %\hspace{-6mm}
\resizebox{170mm}{!}{\rotatebox{0}{\includegraphics{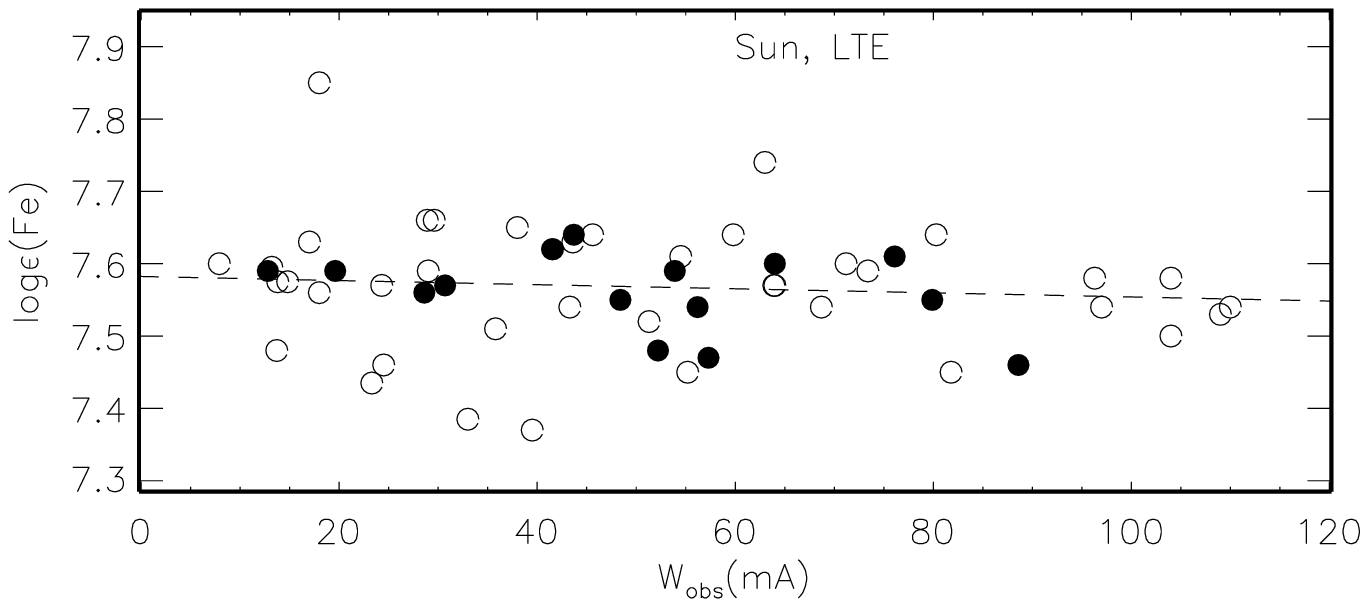}}
\hspace{-8mm}
\rotatebox{0}{\includegraphics{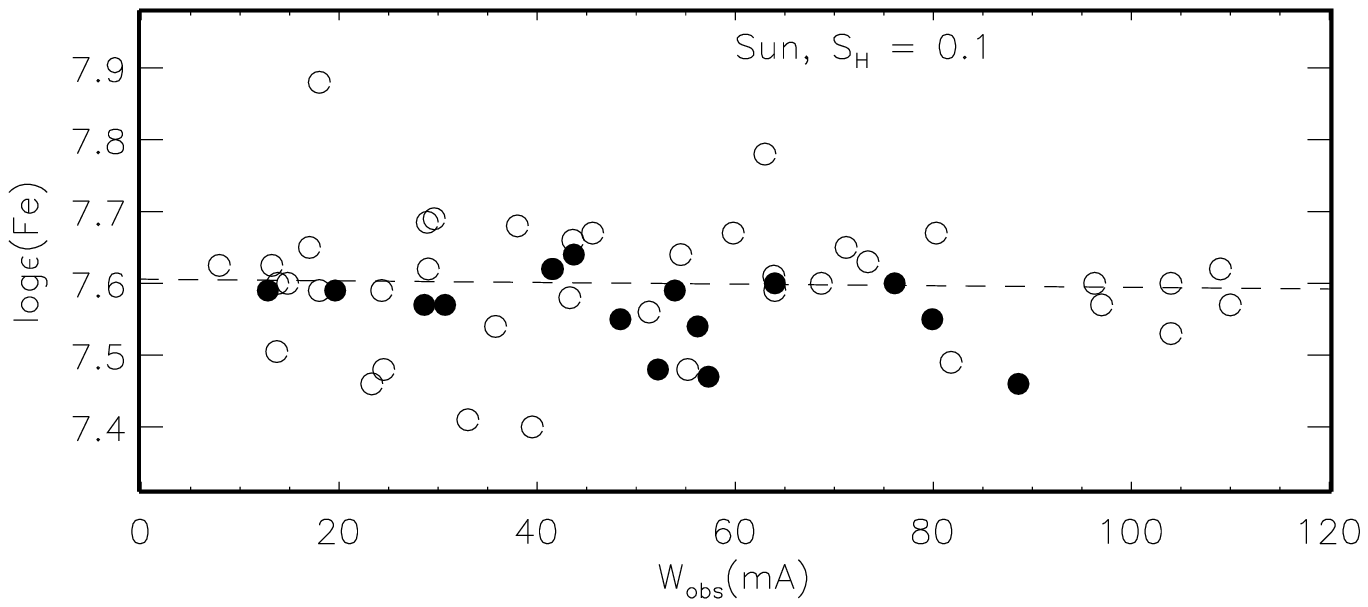}}}}

\vspace{-3mm}
\caption[]{Solar LTE (left column) and non-LTE (\kH\ = 0.1, right column) Fe abundances from the \ion{Fe}{i} (open circles) and \ion{Fe}{ii} (filled circles)
lines plotted as a function of \Eexc\ (top row) and $W_{obs}$ (bottom row). The mean abundances derived from the \ion{Fe}{i} and \ion{Fe}{ii} lines along with their standard deviations are quoted in the top row panels. The dotted line indicates the mean abundance derived from the \ion{Fe}{i} lines and the shaded grey area its statistical error. In the bottom row panels, the dashed line shows the calculated trend of $\eps{\rm Fe I}$ versus $W_{obs}$. Everywhere, $gf-$values indicated in Table\,\ref{linelist} were used.}  \label{Fig:abund_solar}
\end{figure*}

For each line, the element abundance was determined under various
line-formation assumptions: non-LTE with \kH\ = 0, 0.1, 1, and 2, and LTE. The quality of the fits is illustrated in Fig.\,\ref{Fig:line_solar} for two \ion{Fe}{i} lines. The results are shown in Table\,\ref{linelist} (online material) and Fig.\,\ref{Fig:abund_solar}.
To present the absolute abundances from the \ion{Fe}{ii} lines we use $gf-$values from \citet{RU98}, where available, and \citet{Moity83}, just because they provide the mean element abundance consistent with that from the lines of \ion{Fe}{i}.
Compared with their LTE strength \ion{Fe}{i} lines become weaker and \ion{Fe}{ii} lines stronger under non-LTE conditions. The effect is largest for non-LTE with pure electron collisions (\kH\ = 0): the non-LTE abundance correction, $\Delta_{\rm NLTE} = \eps{NLTE}-\eps{LTE}$, ranges between +0.03 and +0.15~dex for various \ion{Fe}{i} lines and between 0.00 and $-0.02$~dex for the \ion{Fe}{ii} lines. As expected, the departures from LTE are weaker in the calculations with \ion{H}{i} collisions taken into account. For the \ion{Fe}{i} lines, $\Delta_{\rm NLTE}$ does not exceed 0.09, 0.04, and 0.03~dex, when \kH\ = 0.1, 1, or 2, respectively. For the \ion{Fe}{ii} lines, the non-LTE abundance correction is smaller than 0.01~dex in absolute values, independent of the \kH\ value. Our calculations showed that the low-excitation lines of \ion{Fe}{i} are subject to stronger deviations from LTE than the higher excitation lines, in agreement with the previous results.
% \citep[see, for example, Table~1 in ][and later non-LTE studies for \ion{Fe}{i}]{Bikmaev90}.
For example, with \kH\ = 0.1, $\Delta_{\rm NLTE}$ ranges between +0.04 and +0.09~dex for six lines with \Eexc\ $<$ 1~eV, while $\Delta_{\rm NLTE}$ = +0.04~dex is the upper limit for the remaining \ion{Fe}{i} lines. As a result, an excitation energy gradient of the non-LTE iron abundances was found to be smaller than that
%, $\eps{\rm Fe I}$\footnote{The subscript in $\eps{\rm Fe I}$ indicates that the iron
%abundance was determined from the neutral species. Likewise, a subscript Fe~II refers to singly ionized lines used in the abundance determination.}  = 7.467 + 0.0221$\times$\Eexc,
for the LTE case. For example, with \kH\ = 0.1, we obtained\footnote{The subscript in $\eps{\rm Fe I}$ indicates that the iron abundance was determined from the neutral species. Likewise, a subscript Fe~II refers to singly ionized lines used in the abundance determination.} $\eps{\rm Fe I}$ = 7.516 + 0.0154$\times$\Eexc (eV).

%{\bf Returning to the subject of the \citet{Blackwell95} and \citet{Holweger1995} controversy, 
We see no significant abundance discrepancies between low-excitation and high-excitation lines. However, the abundance scatter is considerably too high for \Eexc $> 4$~eV, independent of the line-formation assumptions used. For example, the difference between the highest (\ion{Fe}{i} 5517\,\AA) and lowest (\ion{Fe}{i} 5367\,\AA) absolute solar abundance is about 0.5~dex in all the cases. The abundance difference between \ion{Fe}{i} 5662\,\AA\ and \ion{Fe}{i} 5367\,\AA\ with $gf$-values from a common source \citep{fe-OWL91} amounts to about 0.25~dex.
There are essentially two explanations for this result: either the $gf$-values are not reliable or a significant fraction of the iron lines are contaminated by relatively strong (yet undetected) blends or both.
Note that, with an adopted microturbulence velocity of \Vmic\ = 0.9~\kms, the trend of $\eps{\rm Fe I}$ versus $W_{obs}$ is completely removed in the non-LTE calculations.
The \ion{Fe}{i} 5517\,\AA\ line is an obvious outlier in the plots of Fig.\,\ref{Fig:abund_solar}: its abundance is 0.3~dex higher than the mean of the remaining \ion{Fe}{i} lines. The solar mean \ion{Fe}{i} abundances indicated in Table\,\ref{startab} were computed without \ion{Fe}{i} 5517\,\AA\  and \ion{Fe}{i} 5367\,\AA\ taken into account, and the corresponding statistical error is large, $\sigma$ = 0.09~dex.
These results demonstrate how far spectroscopic methods lead us towards accurate and reliable absolute stellar (solar) iron abundance results. Some value around
$\sigma_{\eps{}}$ = 0.1~dex seems to denote the current precision limit.

The difference between the solar mean \ion{Fe}{i} abundances calculated with \kH\ = 0 and 1 amounts to $+0.07$~dex. The corresponding value for \ion{Fe}{ii} is $-0.01$~dex. Since the amplitude of the non-LTE effect is smaller than the combined statistical error of the obtained \ion{Fe}{i} and \ion{Fe}{ii} abundances and smaller than the difference in \ion{Fe}{ii} abundances caused by employing various sources of $gf-$values, {\it we did not endeavor to use the absolute solar Fe abundances determined from two ionization stages to constrain a \kH\ value.}
%of 0.15~dex is larger than of on the abundance difference between \ion{Fe}{i} and \ion{Fe}{ii}. test the non-LTE method.

\subsection{Comparison with other studies}

 During the last decade possibly the most advanced approach to solar abundance determinations is based on ab initio 3D, time-dependent, hydrodynamical model atmospheres.
We compared our 1D LTE and non-LTE solar Fe abundance determinations with the results of 3D calculations performed independently by three scientific groups.
The first 3D solar abundance analysis for Fe
%based on ab initio 3D, time-dependent, hydrodynamical model atmospheres
was made by \citet[][, ANTS2000]{asplund00}. For nine common \ion{Fe}{i} lines, the difference between their 3D+LTE and our 1D+LTE abundances amounts to, on average, $-0.10\pm0.03$~dex.The difference is smaller for the four common \ion{Fe}{ii} lines when using the same $gf-$values, $\Delta$(ANTS2000 - this study) = $-0.04\pm0.01$~dex. In their recent 3D determinations, \citet[][, AGSS2009]{aspl09} applied $gf-$values of the \ion{Fe}{ii} lines from \citet{fe-MB09} and 1D non-LTE correction of $+0.03$~dex for \ion{Fe}{i} and got $\eps{\rm Fe I}$ = 7.52$\pm$0.05  and $\eps{\rm Fe II}$ = 7.50$\pm$0.04 (here, the error bars is the systematic error added in quadrature with the statistical error calculated as the weighted standard error of the mean).
Our 1D non-LTE abundances $\eps{\rm Fe I}$ = 7.56$\pm$0.09 (Table\,\ref{startab}, \kH\ = 0.1) and $\eps{\rm Fe II}$ = 7.47$\pm$0.05 when using $gf-$values from \citet{fe-MB09} (Table\,\ref{Tab:SunAbund}) are consistent within the error bars with the corresponding values of AGSS2009, however, the abundance difference (\ion{Fe}{i} - \ion{Fe}{ii}) = 0.09~dex is larger compared to the 0.02~dex in AGSS2009. This could be due to the difference between 1D and 3D modelling.
Note that updating the method of 3D calculations in AGSS2009 compared to ANTS2000 resulted in the smaller abundance difference (1D - 3D) for \ion{Fe}{i} and
opposite sign of the 3D effect for \ion{Fe}{ii}.

The first non-LTE study for neutral iron that went beyond the 1D analysis was by \citet{ShTB01}. Ignoring inelastic collisions with \ion{H}{i} atoms in their 1.5D+non-LTE calculations, they found a 0.074$\pm$0.03~dex higher solar \ion{Fe}{i} based abundance compared to the LTE value. Our 1D+non-LTE analysis with pure electronic collisions gives for the solar \ion{Fe}{i} lines a very similar mean non-LTE abundance correction of 0.08~dex. However, the absolute abundance is a 0.11~dex higher compared to $\eps{\rm Fe I}$ = 7.50$\pm$0.10 obtained by \citet{ShTB01}.
%, but both values are still consistent within the error bars.

Our 1D \ion{Fe}{ii} based abundance (the non-LTE effects are negligible) is 0.04~dex lower compared to $\eps{\rm Fe II}$ = 7.512$\pm$0.062 obtained by \citet{Caffau2010} from their solar 3D analysis when using $gf-$values from \citet{fe-MB09} in both studies.

 Thus, the difference between our solar iron 1D abundances and the data from the literature based on the 3D calculations does not exceed 0.04~dex when using either the \ion{Fe}{i} or \ion{Fe}{ii} lines. However, the 3D effect is of opposite sign for the two ions, and this may affect analysis of the solar \ion{Fe}{i}/\ion{Fe}{ii} ionization equilibrium.

\begin{table*} %[!t]
\begin{center}
\caption{Characteristics of observed spectra.} \label{Table:obs_data}
\begin{tabular}{lcllccc}
\hline \noalign{\smallskip}
Object & V$^1$ & Telescope /    & Observing run, &\multicolumn{1}{c}{Spectral range}   & $R$ & $S/N$ \\
        & (mag) & spectrograph   & observer       & (\AA)                               &   &     \\
\hline \noalign{\smallskip}
HD\,10700 ($\tau$~Cet)  & 3.50 & 2.2-m / FOCES & Oct. 2001, K. Fuhrmann & 4500 - 6700 & 60\,000 & $\ge 200$ \\
           &      & 2.2-m / FOCES & Oct. 2005, F. Grupp & 4120 - 6700 & 40\,000 & $\ge 200$ \\
HD\,61421 (Procyon) & 0.34 & 2.2-m / FOCES & Feb. 2001, A. Korn & 4200 - 9000 & 60\,000 & $\ge 200$ \\
HD\,84937  & 8.28 & VLT2 / UVES   & ESO UVESPOP & 3300 - 9900            & 80\,000 & $\ge 200$ \\
HD\,102870 ($\beta$~Vir) & 3.61 & 2.2-m / FOCES & May 1997, K. Fuhrmann & 4500 - 6700 & 60\,000 & $\ge 200$ \\
HD\,122563 & 6.20 & VLT2 / UVES   & ESO UVESPOP & 3300 - 9900            & 80\,000 & $\ge 200$ \\
\hline \noalign{\smallskip}
 \multicolumn{7}{l}{ \ $^1$ Visual magnitude from the SIMBAD database.} \\
\end{tabular}
\end{center}
\end{table*}

\section{Stellar sample, observations, and stellar parameters}\label{sect:stars}

Our current sample consists of five stars with the effective temperature and surface gravity determined from methods largely independent of the model atmosphere.
Two of them, HD\,61421 (Procyon) and HD\,102870 ($\beta$~Vir), are close-to-solar metallicity stars. HD\,10700 ($\tau$~Cet) is a mildly metal-deficient star with [Fe/H] $\simeq -0.5$. The two remaining stars are very metal-poor (VMP) with [Fe/H] $< -2$. HD\,84937 represents the hot end of the stars that evolve on time scales comparable to the Galaxy lifetime. HD\,122563, in contrast, is a cool giant. The selected stars are listed in Table\,\ref{Table:obs_data}.

The spectroscopic observations for three of our program stars were carried
out with the fibre-fed \'echelle spectrograph FOCES at the 2.2m telescope
of the Calar Alto Observatory during a number of observing runs, with a spectral resolving power of $R \simeq 60\,000$ (HD\,10700 was also observed at $R \simeq 40\,000$ to get a higher signal-to-noise ($S/N$) spectrum at $\lambda < 4500$\,\AA). Characteristics of the observed spectra are summarized in Table\,\ref{Table:obs_data}. The $S/N$ ratio is 200 or higher in the spectral range $\lambda > 4500$\,\AA, but lower in the blue. All the investigated iron lines lie longwards of 4427\,\AA, and their profile analysis profits from the high spectral resolution and high $S/N$. For HD\,84937 and HD\,122563, we used high-quality observed spectra from the ESO UVESPOP survey \citep{POP03}.

Procyon, $\beta$~Vir, and $\tau$~Cet are among the very few stars for which the whole set of fundamental stellar parameters except metallicity can be determined from (nearly) model-independent methods.
In the most recent study of \citet{Bruntt2010}, the angular diameters of these stars were measured using interfero\-metry. The measured bolometric flux, combined with the angular diameter, implied $\Teff$ = 5383$\pm$47, 6494$\pm$48, and 6012$\pm$64~K for $\tau$~Cet, Procyon, and $\beta$~Vir, respectively. A surface gravity of $\logg$ = 4.54$\pm$0.02, 3.98$\pm$0.02, and 4.13$\pm$0.02, respectively, was computed using the stellar mass and linear radius from Table~2 of \citet{Bruntt2010}. Our adopted stellar parameters of these three stars (Table\,\ref{startab}) were taken from different studies, however, they turned out to be consistent within 1$\sigma$ with the data of \citet{Bruntt2010}.
We employed Procyon's stellar parameters from a careful analysis of \citet{procyon}. The interferometric measurements of the angular diameter and the bolometric flux were used by \citet{betaVir} to derive $\Teff$ and $\logg$ of $\beta$~Vir. For $\tau$~Cet, $\Teff$ and $\logg$ were derived by \citet{tauCet} based on direct interferometric measurements of the angular diameter and stellar evolution calculations.

\begin{table*}
  \begin{center}
\caption{Stellar parameters and obtained iron abundances, [Fe/H],
 for selected stars.}
\label{startab}
\begin{tabular}{llllllrcccc}
\noalign{\smallskip} \hline \noalign{\smallskip}
 Object & $\Teff$, K & $\logg$ & Ref. & \Vmic$^1$ & Ion & N$^2$ & \multicolumn{4}{c}{Iron abundances, [Fe/H]} \\
\cline{8-11}\noalign{\smallskip}
  & & & & & & & LTE & \kH\ = 0 & \kH\ = 0.1 & \kH\ = 1 \\
%\cline{11-12}
%    &         &         &  &  & & & N &   & \\
\noalign{\smallskip} \hline %\noalign{\smallskip}
 \multicolumn{1}{c}{1} & \multicolumn{1}{c}{2} & \multicolumn{1}{c}{3} & \multicolumn{1}{c}{4} & \multicolumn{1}{c}{5} & \multicolumn{1}{c}{6} & \multicolumn{1}{c}{7} & \multicolumn{1}{c}{8} & 9 & 10 & 11  \\
\hline \noalign{\smallskip}
Sun$^3$ & 5777 & 4.44 & & 0.9 & \ion{Fe}{i} & 54 & ~~7.53\scriptsize{$\pm$0.09} & ~~7.61\scriptsize{$\pm$0.10} & ~~7.56\scriptsize{$\pm$0.09} & ~~7.54\scriptsize{$\pm$0.09}  \\
        &      &      & &    & \ion{Fe}{ii}& 18 & ~~7.56\scriptsize{$\pm$0.05} & ~~7.55\scriptsize{$\pm$0.05} & ~~7.56\scriptsize{$\pm$0.05} &  ~~7.56\scriptsize{$\pm$0.05} \\
 HD 10700 & 5377 & 4.53 &                       DTK04& 0.8 & \ion{Fe}{i} & 39 & $-$0.49\scriptsize{$\pm$0.02} & $-$0.44\scriptsize{$\pm$0.03} & $-$0.49\scriptsize{$\pm$0.02} & $-$0.49\scriptsize{$\pm$0.03} \\
  ($\tau$~Cet) &      &      &                      	   &	 & \ion{Fe}{ii}& 13 & $-$0.52\scriptsize{$\pm$0.04} & $-$0.52\scriptsize{$\pm$0.04} & $-$0.52\scriptsize{$\pm$0.04} & $-$0.52\scriptsize{$\pm$0.04} \\
 HD  61421 & 6510\scriptsize{$\pm$49} & 3.96\scriptsize{$\pm$0.02} & AP02 & 1.8 & \ion{Fe}{i} & 45 & $-$0.12\scriptsize{$\pm$0.07} & $-$0.12\scriptsize{$\pm$0.07} & $-$0.10\scriptsize{$\pm$0.07} & $-$0.13\scriptsize{$\pm$0.07} \\
 (Procyon)&      &                            &	   &	 &\ion{Fe}{ii} & 15 & $-$0.03\scriptsize{$\pm$0.05} & $-$0.03\scriptsize{$\pm$0.05} & $-$0.04\scriptsize{$\pm$0.05} & $-$0.03\scriptsize{$\pm$0.05} \\
HD  84937 & 6350\scriptsize{$\pm$37} & 4.09\scriptsize{$\pm$0.05}  & IRFM+Hip & 1.7 & \ion{Fe}{i} & 21 & $-$2.17\scriptsize{$\pm$0.07} & $-$1.96\scriptsize{$\pm$0.06} & $-$2.00\scriptsize{$\pm$0.07} & $-$2.13\scriptsize{$\pm$0.07} \\
       &      &                             &	   &	 & \ion{Fe}{ii}&  9 & $-$2.08\scriptsize{$\pm$0.04} & $-$2.06\scriptsize{$\pm$0.03} & $-$2.08\scriptsize{$\pm$0.04} & $-$2.08\scriptsize{$\pm$0.04} \\
HD 102870 & 6060\scriptsize{$\pm$49} & 4.11\scriptsize{$\pm$0.01}  & NDR09& 1.2 & \ion{Fe}{i} & 39 & ~~0.11\scriptsize{$\pm$0.03} & ~~0.11\scriptsize{$\pm$0.04} & ~~0.13\scriptsize{$\pm$0.04} & ~~0.11\scriptsize{$\pm$0.03} \\
 ($\beta$~Vir) &     &                             &	   &	 & \ion{Fe}{ii}& 13 & ~~0.22\scriptsize{$\pm$0.04} & ~~0.23\scriptsize{$\pm$0.04} & ~~0.22\scriptsize{$\pm$0.04} & ~~0.22\scriptsize{$\pm$0.04} \\
HD 122563 & 4600\scriptsize{$\pm$61} & 1.60\scriptsize{$\pm$0.07}  & IRFM+Hip & 1.95& \ion{Fe}{i} & 35 & $-$2.77\scriptsize{$\pm$0.11} & $-$2.42\scriptsize{$\pm$0.09} & $-$2.61\scriptsize{$\pm$0.09} & $-$2.74\scriptsize{$\pm$0.10} \\
       &      &                             &	   &	 & \ion{Fe}{ii}& 15 & $-$2.56\scriptsize{$\pm$0.07} & $-$2.45\scriptsize{$\pm$0.12} & $-$2.56\scriptsize{$\pm$0.07} & $-$2.56\scriptsize{$\pm$0.07} \\
\noalign{\smallskip} \hline \noalign{\smallskip}
\multicolumn{11}{l}{$ ^1$ Microturbulence velocity is given in \kms.} \\
\multicolumn{11}{l}{$ ^2$ Number of lines used.} \\
\multicolumn{11}{l}{$ ^3$ For the Sun, we present the absolute iron abundances calculated using $gf-$values from Table\,\ref{linelist}.} \\
\multicolumn{11}{l}{ \ \ References to the adopted $\Teff$ and $\logg$: } \\
      \multicolumn{11}{l}{ \ \ DTK04 = \citet{tauCet}, \ \ AP02 = \citet{procyon}, \ \ NDR09 = \citet{betaVir},} \\
      \multicolumn{11}{l}{ \ \ IRFM+Hip = IRFM temperature and {\sc Hipparcos}-parallax based $\logg$.} \\
\end{tabular}
 \end{center}
\end{table*}

For HD\,84937, the infrared flux method (IRFM) temperatures of \citet{alonso96,melendez04,GHB09}, and \citet{CRM10}, $\Teff$ = 6330~K, 6345~K, 6333~K, and 6408~K, respectively, are consistent within 1$\sigma$. Our adopted value, $\Teff$ = 6350~K, is the average of the four. To determine $\logg$, we used an updated {\sc Hipparcos} parallax of $\pi_{Hip}$ = 13.74$\pm$0.78~mas from \citet{Hip_updated} and a mass of 0.8\,${\rm M_\odot}$ derived from the tracks of \citet{vandenberg00}.

For HD\,122563, we adopted $\Teff$(IRFM) = 4600~K recommended by \citet{barbuy} and based on the IRFM determination by \citet{alonso99}. Using the 2MASS photometric system, \citet{GHB09} obtained a 200~K higher IRFM temperature of this star. However, they noted that the 2MASS photometric accuracy is very low for bright giant stars. The gravity was calculated with $\pi_{Hip}$ = 4.22$\pm$0.35~mas from \citet{Hip_updated} and a mass of 0.8\,${\rm M_\odot}$ following \citet{barbuy}. The gravity errors indicated in Table\,\ref{startab} for HD\,84937 and HD\,122563 reflect the uncertainty in the trigonometric parallax.

\section{\ion{Fe}{i} versus \ion{Fe}{ii} in the reference stars}
\label{sect:stellar_iron}

In this section, we derive the \ion{Fe}{i} and \ion{Fe}{ii} abundances in the selected stars under various line-formation assumptions, i.e., non-LTE with \kH\ = 0, 0.1, 1, and 2 and LTE, and we investigate which of them leads to consistent element abundances from both ionization stages.

%\subsection{Abundance analysis}
\subsection{Methodology}

To minimize the effect of the uncertainty in $gf-$values on the final results, we applied a line-by-line differential non-LTE and LTE approach, in the
sense that stellar line abundances were compared with individual abundances of
their Solar counterparts.
Our results for Fe abundances are based on line profile analysis.
In order to compare with
observations, computed synthetic profiles were convolved with a
profile that combines instrumental broadening with a Gaussian
profile, rotational broadening, and broadening by macroturbulence with a radial-tangential profile. Rotational broadening and broadening
by macroturbulence were treated separately only for Procyon and $\beta$~Vir, with $V \sin i$ = 2.6\,\kms\ and 2.5\,\kms, respectively \citep{fuhrmann1998}. For the remaining stars, their overall effect was treated as radial-tangential macroturbulence. The macroturbulence value $\Vmac$ was determined for each star either by \citet{fuhrmann1998,fuhrmann2004} or in our previous
studies \citep{Korn03,ml-nlte-H} from the analysis of an extended list of lines of various chemical species. For a given star, $\Vmac$ was allowed to vary by $\pm$0.3\,\kms\ (1$\sigma$).

For each star, the microturbulence velocity was determined from the requirement that the iron non-LTE abundance derived from \ion{Fe}{i} lines with \kH\ = 0.1 must not depend on the line strength.
We note that, even for the other line-formation scenarios, the slope of the [Fe/H]$_{\rm I}$ - $W_{\rm obs}$ plot is also largely removed with the resulting \Vmic\ value.
% the equivalent width gradient of the \ion{Fe}{i} based \tcol{line} abundances is largely removed.
%Hereafter, the subscript I indicates that the iron abundance was determined from the neutral species. Likewise, a subscript II refers to singly ionized iron lines used in the abundance determination.
%a trend of the element abundance along the line strength was largely removed
%for other line-formation scenarios, too.
%turned out to be as small as for non-LTE with \kH\ = 0.1. For example, with \Vmic\ = 1.95\,\kms\ and \kH\ = 0.1 in the non-LTE calculations for HD\,122563, we achieved [Fe/H] = $-2.609 - 0.150 \cdot 10^{-4} \times W_{obs}$(mA). In LTE and non-LTE with pure electronic collisions, [Fe/H] = $-2.794 + 5.26 \cdot 10^{-4} \times W_{obs}$(mA) and [Fe/H] = $-2.386 - 7.23 \cdot 10^{-4} \times W_{obs}$(mA), respectively.

\begin{figure*}
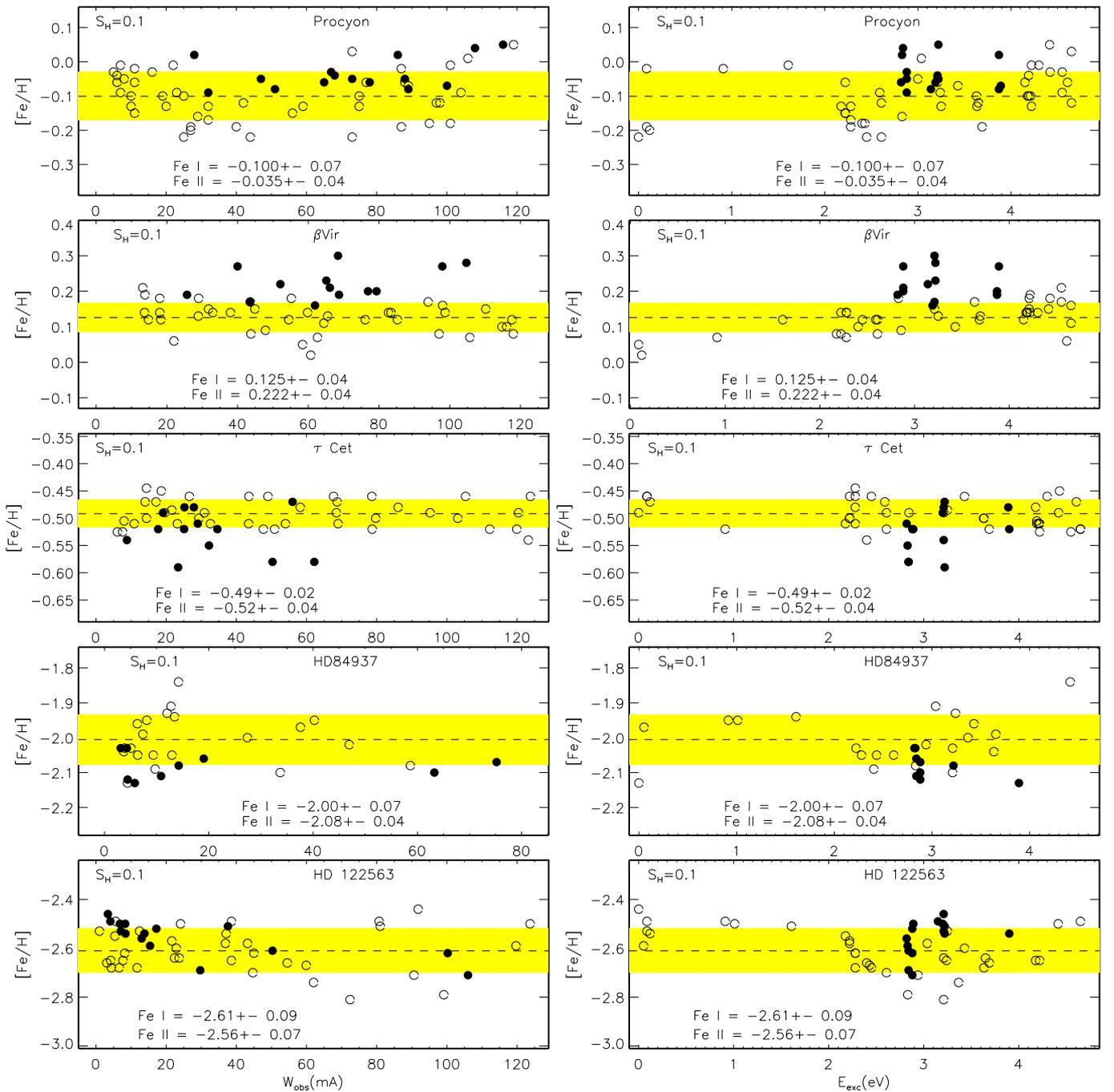

%\vspace*{-3mm}
\hbox{
%\hspace{-6mm}
\resizebox{180mm}{!}{\rotatebox{0}{\includegraphics{fe_wobs_stars_diff.epsi}}
%\hspace{-8mm}
\rotatebox{0}{\includegraphics{fe_exc_stars_diff.epsi}}}}
%\vspace{-5mm}
\caption[]{Non-LTE (\kH\ = 0.1) differential abundances derived from the \ion{Fe}{i} (open circles) and \ion{Fe}{ii} (filled circles) lines in the selected stars as a function of observed equivalent width (left column panels) and excitation energy of the lower level (right column panels). In each panel, the dashed line shows the mean iron abundance determined from the \ion{Fe}{i} lines. See text for more details.}
\label{Fig:fe_wobs}
\end{figure*}

Figure\,\ref{Fig:fe_wobs} shows our final non-LTE (\kH\ = 0.1) abundances from individual lines of two ionization stages in each of the stars investigated. The average \ion{Fe}{i} and \ion{Fe}{ii} abundances obtained in LTE and in non-LTE with \kH\ = 0, 0.1, and 1 are presented in Table~\ref{startab}.
 Having in mind the shortcomings of 1D atmospheric structure modelling, we tend to represent for stars of various metallicity a similar range of the line formation depths. The strongest iron line detected in HD\,122563 has $W_{\rm obs}$ = 123.8\,m\AA, while some of the selected iron lines are very strong in Procyon, $\beta$~Vir, and $\tau$~Cet with $W_{\rm obs}$ up to 250, 305, and 496\,m\AA, respectively. Therefore, we used in stellar abundance and microturbulence velocity analyses only lines with equivalent widths less than 125\,m{\AA}.

\subsection{Non-LTE effects}

\subsubsection{Absolute abundances from \ion{Fe}{ii} lines}

In the stellar parameter range with which we are concerned, the non-LTE effects of the investigated \ion{Fe}{ii} lines are found to be negligible when \ion{H}{i} collisions are taken into account in the SE calculations. For the stars with metal abundances close to solar or mildly deficient, non-LTE with pure electronic collisions leads to a strengthening of the line core for the strongest ($W > 40$\,m\AA) \ion{Fe}{ii} lines because the line source function ($S_{lu} \simeq b_u/b_l\,B_\nu$) drops below the Planck function ($B_\nu$) in the uppermost atmospheric layers. Here, $b_u$ and $b_l$ are the departure coefficients of the upper and lower levels, respectively. The lower levels of the investigated \ion{Fe}{ii} lines are all of even parity with \Eexc\,= 2.8 - 3.9~eV (see Table\,\ref{linelist}, online material), and they keep their thermodynamic equilibrium populations ($b_l = 1$) throughout the atmosphere. The population of the odd parity quartet and sextet terms with \Eexc\,= 5 - 6~eV is controlled by the strong transitions to the ground state a$^6{\rm D}$ and low-excitation states a$^4{\rm F}$, a$^4{\rm D}$, and a$^4{\rm P}$. The line center optical depth of these transitions drops below unity at $\log\tau_{5000} < -2$ (Fig.\,\ref{Fig:bf}), resulting in photon losses in the corresponding lines and underpopulation of the upper levels ($b_u < 1$).
For Procyon, $\beta$~Vir, and $\tau$~Cet, the non-LTE abundance correction is negative and does not exceed 0.02~dex in absolute value. The departures from LTE are negligible for the weaker lines. Non-LTE with \kH\ = 0 results in the opposite effect for the VMP stars. This can be understood as follows. In the atmosphere with [Fe/H] = $-2.5$, the line wing optical depth drops below 1 far inside the atmosphere even for the strongest \ion{Fe}{ii} transitions arising from the ground and low-excitation states to the odd parity levels with \Eexc\,= 5 - 6~eV. As a result, they
are pumped by the ultraviolet $J_\nu - B_\nu(T_e)$ excess radiation and produce enhanced excitation of the odd parity levels in the layers with $\log\tau_{5000}$ between +0.3 and $-1$ (Fig.\,\ref{Fig:bf}).
Here, the term "pumped'' (or pumping) is used following \citet{NaK}. In the upper layers, $\log\tau_{5000} < -1$, where the line center optical depth of the UV transitions drops below 1, enhanced excitation changes with photon losses resulting in a decrease of the departure coefficients of the odd parity levels. The investigated \ion{Fe}{ii} lines are formed in the atmospheres of HD\,84937 and HD\,122563 inside $\log\tau_{5000} = -1.4$, where, for every line, the source function exceeds the Planck function, resulting in weakening the line relative to its LTE strength.
%weakened because in the line formation layers ($-1.4 < \log\tau_{5000} < %0$) due to $b_u/b_l > 1$.
The average non-LTE abundance correction amounts to +0.01~dex and +0.10~dex, respectively.
%The UV line pumping produces a strongly enhanced excitation of the %\ion{Fe}{ii} levels with \Eexc\ $>$ 7.3~eV in the line formation layers %of the metal-poor models as can be seen in Fig.\,\ref{Fig:bf} (right %column).

\subsubsection{Absolute abundances from \ion{Fe}{i} lines}

In all cases, non-LTE leads to a weakening of the \ion{Fe}{i} lines. In close-to-solar metallicity models, this is mainly due to overionization. For each line, the source function is quite similar to the Planck function because most levels behave similarly, as can be seen in Fig.\,\ref{Fig:bf} (left top panel). At the other end of the metallicity range, the energy levels become weakly coupled far inside the VMP atmospheres due to
deficient collisions (Fig.\,\ref{Fig:bf}, right top panel). At the
depths where the \ion{Fe}{i} lines are formed, the upper levels are all
depleted to a lesser extent relative to their LTE populations than
are the lower levels. The lines are weaker relative to their LTE
strengths not only due to the general overionization but also due
to $b_u/b_l > 1$ resulting in $S_{lu} > B_\nu$ and the depleted line absorption.
 The non-LTE effects are the strongest when \ion{H}{i} collisions are neglected in the SE calculations. On average, $\Delta_{\rm NLTE}$(\kH\ = 0) = +0.05~dex and +0.06~dex for Procyon and $\beta$~Vir and increases with decreasing metallicity to +0.13~dex at [Fe/H] = $-0.5$ ($\tau$~Cet), +0.26~dex at [Fe/H] = $-2.1$ (HD\,84937), and  +0.43~dex at [Fe/H] = $-2.5$ (HD\,122563). With \ion{H}{i} collisions taken into account, the effect on the iron abundance determination is substantially weaker. For example, with \kH\ = 0.1, $\Delta_{\rm NLTE}$ = 0.00, +0.04, 0.03, 0.19, and 0.19~dex for the same sequence of stars, respectively.

\subsubsection{Differential abundances from \ion{Fe}{i} lines}

For stellar differential abundances, the non-LTE correction can be introduced as
$[\Delta_{\rm NLTE}]$ = [Fe/H]$_{\rm NLTE}$ -- [Fe/H]$_{\rm LTE}$.
Since the departures from LTE for the solar \ion{Fe}{i} lines are small in the calculations with \kH\ $\ne 0$, the corresponding $[\Delta_{\rm NLTE}]$ and $\Delta_{\rm NLTE}$ values turned out to be close together. For example,
$[\Delta_{\rm NLTE}]$(\kH\ = 0.1) = +0.02, 0.02, 0.00, 0.17, and 0.16~dex for Procyon, $\beta$~Vir, $\tau$~Cet, HD\,84937, and HD\,122563, respectively. In the \kH\ = 0 case, the non-LTE differential abundance correction is approximately 0.08~dex lower than the corresponding $\Delta_{\rm NLTE}$ value.

\subsubsection{\ion{Fe}{i} excitation equilibrium}

For the two VMP stars of our sample, non-LTE leads to the shallower excitation gradient of the \ion{Fe}{i} line abundances compared to that obtained under the LTE assumption. For example, with \kH\ = 0.1 (the right column panels of Fig.\,\ref{Fig:fe_wobs}), $d{\rm [Fe/H]}/d$\Eexc\ = +0.013~dex/eV for HD\,84937 and $-0.030$~dex/eV for HD\,122563, while the corresponding LTE values are $-0.019$ and $-0.054$~dex/eV, respectively. However, the non-LTE gradient obtained for HD\,122563 is still steep compared to other stars of the sample. Our test calculations show that it could be substantially reduced, down to $d{\rm [Fe/H]}/d$\Eexc\ $= -0.008$~dex/eV, with an effective temperature revised downward by 80~K (1$\sigma$). No significant change in the excitation slope between non-LTE and LTE was found for Procyon, $\beta$~Vir, and $\tau$~Cet, because the departures from LTE for \ion{Fe}{i} are small in these stars. The gradient is small, at the level of $-0.005$ to $-0.007$~dex/eV, for $\tau$~Cet. For Procyon and $\beta$~Vir, $d{\rm [Fe/H]}/d$\Eexc\ $\simeq 0.02$~dex/eV. We caution against determining stellar effective temperatures of metal-poor stars from the \ion{Fe}{i} LTE excitation equilibrium.
%  Since the excitation gradient cannot be removed completely even in non-LTE analysis,

\subsection{Empirically constraining the efficiency of \ion{H}{i} collisions}

As can be seen from Table~\ref{startab}, the behavior of the difference between \ion{Fe}{i} and \ion{Fe}{ii} abundances, $\Delta$(\ion{Fe}{ii} - \ion{Fe}{i}) = [Fe/H]$_{\rm II}$ -- [Fe/H]$_{\rm I}$, within various line-formation scenarios is different for stars of different metal abundance.

A disparity between the neutral and singly-ionized iron is found in Procyon and in $\beta$~Vir,
independent of either LTE or non-LTE. The least imbalance achieved in non-LTE with \kH\ = 0.1 amounts to 0.06~dex and 0.09~dex, respectively. Very similar results for \ion{Fe}{i}/\ion{Fe}{ii} in Procyon were obtained earlier by \citet{Korn03}. For the same star, a similar problem was uncovered also for \ion{Ca}{i}/\ion{Ca}{ii} \citep{mash_ca}. Such a disparity can be caused by the uncertainties either in atmosphere and line-formation modelling or in stellar parameters. The three-dimensional LTE simulations for Procyon \citep{procyon} showed that weak lines ($W_\lambda \le$ 50\,m\AA) of both \ion{Fe}{i} and \ion{Fe}{ii} are weakened compared to a classical 1D analysis, such that the derived Fe abundance of this star increases by 0.05\,dex and 0.04\,dex, respectively. Thus, LTE+3D modelling does not master \ion{Fe}{i}/\ion{Fe}{ii} in Procyon. The solution of the non-LTE+3D problem for iron is still a challenge for stellar spectra modelling (see Sect.\,\ref{sect:uncertainty}).
As discussed by \citet{Korn03}, to remove a discrepancy of 0.07~dex between \ion{Fe}{ii} and \ion{Fe}{i} in Procyon, one would require either a gravity of $\logg$ = 3.81, many standard deviations away from the astrometric result, or a temperature of $\Teff$ = 6600~K, almost 2$\sigma$ higher than the temperature based on bolometric flux. Our test calculations with an 80~K upward revised $\Teff$ achieved consistent \ion{Fe}{i} and \ion{Fe}{ii} abundances of this object (Table\,\ref{Tab:uncertaity}).
%This study was not aimed to revise stellar parameters of the investigated stars. We rather checked whether the uncertainty in line-formation modelling can be responsible, at least in part, for the abundance inconsistencies.
%However, $\Delta$( - ) weakly depends on the used line-formation assumptions, and none of these stars can, therefore, serve as a calibrator for \kH. The more worrying point is inconsistent Fe abundances obtained from two ionization stages in the stars with the precise stellar parameters, when applying a line-by-line differential approach and the refined method of line-formation modelling. Though it should be noted that $\Delta$(\ion{Fe}{i} - \ion{Fe}{ii}) does not exceed 2$\sigma$, in all cases. was found earlier
% In , we investigate the effect of the uncertainties in the used atomic data on the derived iron non-LTE abundances in Procyon.

We find that non-LTE with \kH\ $\ge 0.1$ is as good as LTE to achieve the ionization equilibrium between \ion{Fe}{i} and \ion{Fe}{ii} in $\tau$\,Cet. Yet non-LTE with pure electronic collisions cannot be excluded completely, because an abundance difference of $\Delta$(\ion{Fe}{ii} - \ion{Fe}{i}, \kH\ = 0) = $-0.08$ is only by a factor of 1.5 larger than its error bars.

We come to the conclusion that non-LTE is a must when analyzing very metal-poor stellar spectra, whether hot or cool. LTE leads to a discrepancy in abundances between \ion{Fe}{ii} and \ion{Fe}{i} of 0.09~dex for HD\,84937, which is still at the level of 1$\sigma$, and 0.21~dex for HD\,122563. As can be seen from Table~\ref{startab}, $\Delta$(\ion{Fe}{ii} - \ion{Fe}{i}) should approach 0 in non-LTE with a \kH\ value of between 0.1 and 1 for HD\,84937 and between 0 and 0.1 for HD\,122563. With \kH\ = 0.1, $\Delta$(\ion{Fe}{ii} - \ion{Fe}{i}) = $-0.08$~dex for the first of them and $+0.05$~dex for the second one. Most F5 - K0 type stars in the super-solar metallicity down to [Fe/H] = $-2.5$ domain have stellar parameters $\Teff$ and $\log g$ in between those for HD\,84937 (6350/4.09) and HD\,122563 (4600/1.60). Therefore, the uncertainty in the estimated \kH\ value is expected to result in an abundance error of no more than 0.08~dex in the non-LTE calculations for \ion{Fe}{i} in such a type of stars.
Note that this abundance error decreases towards higher metallicity. For comparison, an abundance error due to using the LTE assumption can be as large as 0.2~dex.

\section{Uncertainties in the iron non-LTE abundances}\label{sect:uncertainty}

To assess the influence of crucial atomic data and stellar parameters on the accuracy of iron
non-LTE abundances, test calculations were performed for Procyon, HD\,84937, and HD\,122563. For each parameter or atomic model that we varied, we derived stellar Fe abundances from the \ion{Fe}{i} and \ion{Fe}{ii} lines and then the average values. The results of the tests are summarized in Table\,\ref{Tab:uncertaity}.

\begin{table*}
\caption{Uncertainties in the non-LTE analysis of \ion{Fe}{i} and \ion{Fe}{ii} in selected stars.}
\label{Tab:uncertaity} \tabcolsep3.5mm
\begin{center}
\begin{tabular}{lrrrrrrrr}
\hline\noalign{\smallskip}
 & \multicolumn{8}{c}{Changes in non-LTE abundances, [Fe/H], relative} \\
 & \multicolumn{8}{c}{to those in Col.~10 (\kH\ = 0.1) of Table \ref{startab}} \\
\noalign{\smallskip} \cline{2-9} \noalign{\smallskip}
Input parameter & \multicolumn{2}{c}{~~HD\,61421} & \ \ & \multicolumn{2}{c}{HD\,84937} & \ \ & \multicolumn{2}{c}{HD\,122563} \\
 & \ion{Fe}{i} &  \ion{Fe}{ii} & & \ion{Fe}{i} &  \ion{Fe}{ii} & & \ion{Fe}{i} & \ion{Fe}{ii} \\
\noalign{\smallskip} \hline \noalign{\smallskip}
\ion{H}{i} collisions: & \multicolumn{8}{c}{ } \\
 \ \ \kH\ = 0   & $-0.02$ & $0.01$ & & $+0.04$ & $+0.02$ & & $+0.19$ & $+0.11$ \\
 \ \ \kH\ = 1   & $-0.03$ & $0.01$ & & $-0.13$ & $0.00$ & & $-0.13$ & $0.00$ \\
 \ \ \kH\ = 2   & $-0.03$ & $0.00$ & &$-0.14$ & $0.00$ & & $-0.15$ & $0.00$ \\
 \ \ \kH\ = 0.1 / 0 for & $0.00$ & $0.00$ & & $+0.06$ & $0.00$ & & $+0.08$ & $+0.03$ \\
 \ \ \ allowed / forbidden transitions & \multicolumn{8}{c}{ } \\
Reduced model atom, \kH\ = 0.1 & $+0.01$ & $+0.01$ & & $+0.02$ & $+0.01$ & & $+0.07$ & $0.00$ \\
Stellar parameters: & \multicolumn{8}{c}{ } \\
 \ \  $\Delta \Teff = -80$ K &- &- & & - &- & & $-0.14$ &  $+0.01$ \\
 \ \  $\Delta \Teff = +80$ K & $+0.06$ & $0.00$ & & $+0.06$ & $+0.03$ & & - &-  \\
 \ \  $\Delta \logg = -0.1$  & - & - & & $0.00$ & $-0.03$ & & $+0.02$ &  $-0.04$ \\
 \ \  $\Delta$\Vmic\ $= -0.1$~\kms & 0.02 & 0.04 & & 0.00 & 0.01 & & 0.02 &  0.01 \\
\noalign{\smallskip}\hline
\end{tabular}
\end{center}
\end{table*} %

\subsection{Hydrogenic collisions}

The effect of including hydrogenic collisions in our SE calculations is shown in the first four lines of Table\,\ref{Tab:uncertaity}. We discussed above how the \ion{Fe}{i} and \ion{Fe}{ii} abundances vary depending on the \kH\ value. Here, we concentrate on the effect of using the atomic model, where \ion{H}{i} collisions were taken into account for the allowed transitions, but were neglected for the forbidden transitions. For a differential analysis of stellar spectra, the calculations were performed with \kH\ = 0, 0.1, and 1 not only for the three tested stars, but also for the Sun.
As expected, neglecting hydrogenic collisions for the forbidden transitions leads to substantially stronger departures from LTE for \ion{Fe}{i} in the VMP stars. With the modified model atom, {\it no single} \kH\ value leading to consistent \ion{Fe}{i} and \ion{Fe}{ii} abundances in both HD\,84937 and HD\,122563 was found.

\subsection{Completeness of the model atom}

For the Sun and the reference stars, calculations were also performed with the reduced atomic model, where all the predicted high-excitation levels of \ion{Fe}{i} were removed. As shown in Sect.\,\ref{sect:atom_compl}, the use of the reduced model atom leads to a stronger overionization of \ion{Fe}{i} compared to that calculated with our final model atom. As a result, the \ion{Fe}{i} based non-LTE abundances obtained with the reduced model atom are higher than the corresponding values in Table \ref{startab}. For the Sun, the abundance difference amounts to 0.08~dex with \kH\ = 0.1.
The changes in stellar differential abundances are smaller (Table\,\ref{Tab:uncertaity}). An exception is HD\,122563, where the \ion{Fe}{i} mean abundance changes by +0.07~dex. We {\it caution against applying an incomplete model atom} of neutral iron or any other minority species to a non-LTE analysis of stellar spectra.

\subsection{Stellar parameters}

We investigated how the uncertainties in stellar parameters influence our choice of the \kH\ scaling factor.
%The direction of changes in $\Teff$ and $\logg$ was chosen to reduce the difference in iron LTE abundances between two ionization stages. Therefore,
 For both VMP stars of our sample, the test calculations were performed with a 0.1~dex (more than 1$\sigma$) lower gravity and an 80~K (approximately 2$\sigma$) revised effective temperature, upwards for HD\,84937 and downwards for HD\,122563.
%For HD\,122563, we also checked how the excitation gradient of the \ion{Fe}{i} based abundances varies with reducing $\Teff$ by 80~K (more than 1$\sigma$).
The results of these tests can be summarized as follows.
\begin{itemize}
\item The variation in $\Teff$, $\logg$, and \Vmic\ within the error bars does not help to achieve the \ion{Fe}{i}/\ion{Fe}{ii} ionization equilibrium in HD\,84937 and HD\,122563 under the LTE assumption.
\item Reducing the gravity by 0.1~dex does not change the excitation gradient of the \ion{Fe}{i} based abundances in each of these two stars. A relatively steep trend of [Fe/H]$_{\rm I}$ versus \Eexc\ obtained in HD\,122563 can essentially be removed with a downward revision of the effective temperature by 80~K.
\item The uncertainties in stellar parameters do not influence our conclusion that there is a need for inelastic collisions with hydrogen atoms (or other thermalizing collisional processes not involving electrons) to establish the SE of iron in the atmospheres of metal-poor stars. With the downward revised $\Teff$ or $\logg$, HD\,122563 favors a low efficiency of \ion{H}{i} collisions with \kH\ = 0 and 0.1, respectively. With the higher $\Teff$ or the lower gravity of HD\,84937, the difference between [Fe/H]$_{\rm I}$ and [Fe/H]$_{\rm II}$ is removed with \kH\ $\simeq 1$.
\end{itemize}

\subsection{Stellar atmosphere and line formation modelling}

Only two non-LTE studies for iron exist in the literature that go beyond the 1D analysis. \citet{ShTB01} and \citet{sh_Fe_2005} performed 1.5D non-LTE calculations in 3D model atmospheres of the Sun and the metal-poor subgiant HD\,140283. They found that the departures from LTE are stronger in 3D than in 1D model and gave a 3D non-LTE abundance correction of $\Delta_{\rm 3D,NLTE} = \eps{3D,NLTE} - \eps{3D,LTE}$ = +0.07~dex for the Sun and about +0.9~dex for HD\,140283. \citet{sh_Fe_2005} have demonstrated that the ionization equilibrium between \ion{Fe}{i} and \ion{Fe}{ii} in HD\,140283 cannot be achieved with $\Teff$ = 5700~K, $\logg$ = 3.7, and [Fe/H] $= -2.5$, and they obtained surprisingly large statistical errors of $\sigma_{\eps{}}$ = 0.19~dex and 0.48~dex for the abundances determined from the \ion{Fe}{i} and \ion{Fe}{ii} lines, respectively. They argued that adopting $\Teff \simeq$ 5600~K and [Fe/H] $\simeq -2.0$ substantially reduces the discrepancies between the two ionization stages. We note that this effective temperature is in conflict with the recently improved IRFM temperature of this star, either $\Teff$ = 5755~K \citep{GHB09} or $\Teff$ = 5777~K \citep{CRM10}. An explanation for this result can be the use of an incomplete model atom of \ion{Fe}{i} and ignoring inelastic collisions with \ion{H}{i} atoms. The \ion{Fe}{i} term diagram applied by \citet{sh_Fe_2005} is, in fact, complete up to \Eexc\,=5.72~eV. But, at higher energies, between \Eexc\,=5.72~eV and 7.0~eV, it only contains about 50\,\%\ of the terms identified by \citet{Nave1994}.
The non-LTE+3D problem for iron still awaits a more satisfactory solution.

\section{Conclusions} \label{conclusion}

In this study, a comprehensive model atom for neutral and singly-ionized iron was built up using atomic data for the energy levels and transition probabilities from  laboratory measurements and theoretical predictions. With a fairly complete model atom for \ion{Fe}{i}, the calculated statistical equilibrium of iron changed substantially by achieving close collisional coupling of
the \ion{Fe}{i} levels near the continuum to the ground state of \ion{Fe}{ii}. There is no need anymore for the enforced upper level thermalization procedure that was applied in the previous non-LTE analyses \citep{Gehren2001a,Korn03,Colletal05}.

Non-LTE line formation for \ion{Fe}{i} and \ion{Fe}{ii} lines was considered in 1D model atmospheres of the Sun and five reference stars with reliable stellar parameters, which cover a broad range of effective temperatures between 4600~K and 6500~K, gravities between $\logg$ = 1.60 and 4.53, and metallicities between [Fe/H] = $-2.5$ and $+0.1$. We found that the departures from LTE are negligible for the \ion{Fe}{ii} lines over the whole stellar parameter range considered. For \ion{Fe}{i}, the non-LTE effects on the abundances are expected to be small for stars with solar-type metallicities such as the Sun, Procyon, and $\beta$~Vir, and for mildly metal-deficient stars such as $\tau$~Cet: a non-LTE correction is at the level of $+0.1$~dex in non-LTE with pure electronic collisions and of a few hundredths, when inelastic collisions with hydrogen atoms are taken into account in the SE calculations with \kH\ $\ge 0.1$.

From a differential line-by-line analysis of stellar spectra we found that the iron ionization equilibrium is not fulfilled in Procyon and $\beta$~Vir at their fundamental parameters, 6510/3.96 and 6060/4.11, respectively, independent of either LTE or non-LTE. For Procyon, an upward revision of its temperature by 80~K removes the obtained abundance imbalance.

In contrast, consistent iron abundances from both ionization stages were obtained in $\tau$~Cet at its given stellar parameters in both non-LTE with \kH\ $\ge 0.1$ and in LTE.

Significant departures from LTE for \ion{Fe}{i} were found in the two VMP stars of our sample, HD\,84937 and HD\,122563.
Our results indicate the need for a thermalizing process not involving electrons in their atmospheres.
 Since there are no accurate theoretical considerations of appropriate processes for iron, we simulate an additional source of thermalization in the atmospheres of cool stars by parametrized \ion{H}{i} collisions. 
Close inspection of the ionization equilibrium between \ion{Fe}{i} and \ion{Fe}{ii} in HD\,84937 and HD\,122563 leads us to choose {\it a scaling factor of 0.1} to the formula of
\citet{Steenbock1984} for calculating hydrogenic collisions. 
The uncertainty in the estimated \kH\ value is expected to result in abundance error of no more than 0.08~dex in the non-LTE calculations for \ion{Fe}{i} in the F5 - K0 type stars in the super-solar metallicity down to [Fe/H] = $-2.5$ domain.
 However, the situation can be significantly worse for extremely and ultra-metal-poor stars. Exactly for such objects, the determination of the surface gravity relies in most cases on the analysis of the \ion{Fe}{i}/\ion{Fe}{ii} ionization equilibrium. Theoretical studies are urgently needed to evaluate the cross-sections of inelastic
collisions of \ion{Fe}{i} with \ion{H}{i} atoms and to search for and evaluate other types of thermalizing processes.

For the Sun, the use of \kH\ = 0.1
leads to an average \ion{Fe}{i} non-LTE correction of 0.03~dex and a mean \ion{Fe}{i} based abundance of 7.56$\pm$0.09. A mean solar abundance derived from \ion{Fe}{ii} lines varies between 7.41$\pm$0.11 and 7.56$\pm$0.05 depending on the source of $gf-$values. A statistical error of 0.09 -- 0.11~dex is uncomfortably high for the Sun. It is, most probably, due to the uncertainty in $gf-$values and, in part, van der Waals damping constants. The problem of oscillator strengths of the Fe lines and their influence on the derived solar iron abundance has been debated for decades \citep{Blackwell95,Holweger1995,Kostik1996,Caffau2010}. Stellar astrophysics needs accurate atomic data for an extended list of the iron lines, which could be measured in stars of very different metallicities.
We therefore call on laboratory atomic spectroscopists for further efforts to improve $gf-$values of the \ion{Fe}{i} and \ion{Fe}{ii} lines used in abundance analysis.

Using our carefully calibrated model of iron, cool stars over a broad range of metallicities encountered in the Galaxy can now be analyzed in a homogeneous way to derive their iron abundance and gravity without resorting to trigonometric parallaxes. The dependence of departures from LTE on stellar parameters and an application of the non-LTE technique to the known ultra-metal-poor stars ([Fe/H] $< -4.5$)
will be presented in a forthcoming paper.

\begin{acknowledgements}
  The authors thank Manuel Bautista and Tatyana Ryabchikova for help with collecting the atomic data and Nicolas Grevesse for useful discussion of the accuracy of $gf-$values for the iron lines. This research was supported by the Russian Foundation for Basic Research (08-02-92203-GFEN-a and 08-02-00469-a), the Russian Federal Agency on Science and Innovation (02.740.11.0247), the Deutsche Forschungsgemeinschaft (GE 490/34.1), and the National Natural Science Foundation of China (No. 10973016). A.J.K. acknowldeges support by the Swedish Reseach Council and the Swedish National Space Board. We also thank the anonymous (second) referee for valuable suggestions and comments.
  We made use of the NIST, SIMBAD, and VALD databases.

\end{acknowledgements}

\Online

\longtab{5}{
\begin{longtable}[]{ccrcccccccccl}
\caption{\label{linelist} Line data and iron LTE and non-LTE abundances from an analysis of the solar spectrum. }  \\
\hline \noalign{\smallskip}
 $\lambda$ & Transition & Mult & $E_{exc}$ & $\log gf$ & $\log C_6$ & \multicolumn{5}{c}{$\eps{}$} & $\Vmac$ & Ref.$ ^1$ \\
\cline{7-11}
  (\AA)    &            &      &  (eV)     &           &            & LTE  & \kH\ = 0 & 0.1 & 1 & 2 & 0.1 & \\
\hline
 1 & 2 & \multicolumn{1}{c}{3} & 4 & \multicolumn{1}{c}{5} & \multicolumn{1}{c}{6} & 7 & \multicolumn{1}{c}{8} & \multicolumn{1}{c}{9} & \multicolumn{1}{c}{10} & \multicolumn{1}{c}{11} & \multicolumn{1}{c}{12} & \multicolumn{1}{c}{13}\\
\hline \noalign{\smallskip}
\endfirsthead
\caption{continued.}\\
\hline \noalign{\smallskip}
 $\lambda$ & Transition & Mult & $E_{exc}$ & $\log gf$ & $\log C_6$ & \multicolumn{5}{c}{$\eps{}$} & $\Vmac$ & Ref.$ ^1$ \\
\cline{7-11}
  (\AA)    &            &      &  (eV)     &           &            & LTE  & \kH\ = 0 & 0.1 & 1 & 2 & 0.1 & \\
\hline
 1 & 2 & \multicolumn{1}{c}{3} & 4 & \multicolumn{1}{c}{5} & \multicolumn{1}{c}{6} & 7 & \multicolumn{1}{c}{8} & \multicolumn{1}{c}{9} & \multicolumn{1}{c}{10} & \multicolumn{1}{c}{11} & \multicolumn{1}{c}{12} & \multicolumn{1}{c}{13} \\
\hline \noalign{\smallskip}
\endhead
\hline
\endfoot
\hline
%\hline
\endlastfoot
 \multicolumn{13}{c}{ \ion{Fe}{i} lines} \\
     5166.28 & \eu{a}{5}{D}{}{4} - \eu{z}{7}{D}{\circ}{5} &	 1  &  0.00 & -4.20 &-31.93 &  7.53 &  7.68 &  7.62 &  7.57 &  7.56 & 2.6& BIP79 \\
     5247.06 & \eu{a}{5}{D}{}{2} - \eu{z}{7}{D}{\circ}{3} &	 1  &  0.09 & -4.95 &-31.92 &  7.54 &  7.64 &  7.60 &  7.56 &  7.55 & 2.7& BIP79 \\
     5250.21 & \eu{a}{5}{D}{}{0} - \eu{z}{7}{D}{\circ}{1} &	 1  &  0.12 & -4.94 &-31.90 &  7.60 &  7.75 &  7.65 &  7.61 &  7.61 & 2.9& BIP79 \\
     4427.31 & \eu{a}{5}{D}{}{3} - \eu{z}{7}{F}{\circ}{4} &	 2  &  0.05 & -2.92 &-31.86 &  7.42 &  7.52 &  7.47 &  7.44 &  7.44 & 2.2& OWL91   \\
     4445.47 & \eu{a}{5}{D}{}{2} - \eu{z}{7}{F}{\circ}{2} &	 2  &  0.09 & -5.44 &-31.86 &  7.54 &  7.60 &  7.58 &  7.55 &  7.55 & 2.7& BIP79 \\
     5434.53 & \eu{a}{5}{F}{}{1} - \eu{z}{5}{D}{\circ}{0} &	15  &  1.01 & -2.12 &-31.74 &  7.41 &  7.46 &  7.45 &  7.44 &  7.43 & 2.6& FMW88 \\
     4994.13 & \eu{a}{5}{F}{}{4} - \eu{z}{5}{F}{\circ}{3} &	16  &  0.91 & -2.96 &-31.71 &  7.38 &  7.45 &  7.43 &  7.41 &  7.41 & 2.6& BKK91  \\
     5216.27 & \eu{a}{3}{F}{}{2} - \eu{z}{3}{F}{\circ}{2} &	36  &  1.61 & -2.15 &-31.52 &  7.41 &  7.44 &  7.44 &  7.43 &  7.43 & 2.8& FMW88 \\
     6151.62 & \eu{a}{5}{P}{}{3} - \eu{y}{5}{D}{\circ}{2} &	62  &  2.18 & -3.30 &-31.58 &  7.52 &  7.63 &  7.56 &  7.53 &  7.53 & 2.9& BPS82a \\
     6213.43 & \eu{a}{5}{P}{}{1} - \eu{y}{5}{D}{\circ}{1} &	62  &  2.22 & -2.48 &-31.58 &  7.45 &  7.56 &  7.49 &  7.47 &  7.47 & 2.8& OWL91  \\
     6082.71 & \eu{a}{5}{P}{}{1} - \eu{z}{3}{P}{\circ}{1} &	64  &  2.22 & -3.57 &-31.74 &  7.51 &  7.61 &  7.54 &  7.52 &  7.51 & 3.0& BPS82a \\
     5198.72 & \eu{a}{5}{P}{}{1} - \eu{y}{5}{P}{\circ}{2} &	66  &  2.22 & -2.14 &-31.32 &  7.50 &  7.59 &  7.53 &  7.52 &  7.52 & 2.7& BPS82a \\
     6481.88 & \eu{a}{3}{P}{}{2} - \eu{y}{5}{D}{\circ}{2} &   109  &  2.28 & -2.98 &-31.44 &  7.57  &  7.69 &  7.61 &  7.58 &  7.58 & 2.8& BPS82a \\
     6608.03 & \eu{a}{3}{P}{}{2} - \eu{y}{5}{D}{\circ}{3} &   109  &   2.28 & -4.03 &-31.61 &  7.56 &  7.66 &  7.59 &  7.57 &  7.56 & 3.0& FMW88 \\
     6421.35 & \eu{a}{3}{P}{}{2} - \eu{z}{3}{P}{\circ}{2} &   111  &   2.28 & -2.03 &-31.80 &  7.54 &  7.60 &  7.57 &  7.55 &  7.55 & 2.9& BPS82a \\
     4574.72 & \eu{a}{3}{P}{}{2} - \eu{x}{5}{D}{\circ}{2} &   115  &   2.28 & -2.97 &-31.09 &  7.64 &  7.74 &  7.67 &  7.65 &  7.65 & 3.0& FMW88 \\
     6393.61 & \eu{a}{3}{H}{}{5} - \eu{z}{5}{G}{\circ}{4} &   168  &   2.43 & -1.43 &-31.53 &  7.46 &  7.54 &  7.49 &  7.47 &  7.47 & 2.8& BKK91 \\
     6252.55 & \eu{a}{3}{H}{}{6} - \eu{z}{3}{G}{\circ}{5} &   169  &   2.40 & -1.69 &-31.52 &  7.58 &  7.65 &  7.61 &  7.59 &  7.59 & 2.8& BPS82a \\
     5916.25 & \eu{a}{3}{H}{}{4} - \eu{y}{3}{F}{\circ}{4} &   170  &   2.45 & -2.99 &-31.45 &  7.61 &  7.72 &  7.64 &  7.62 &  7.62 & 2.9& BPS82a \\
     6065.49 & \eu{b}{3}{F}{}{2} - \eu{y}{3}{F}{\circ}{2} &   207  &   2.61 & -1.53 &-31.41 &  7.57 &  7.63 &  7.59 &  7.58 &  7.58 & 2.8& BPS82b \\
     6200.32 & \eu{b}{3}{F}{}{2} - \eu{y}{3}{F}{\circ}{3} &   207  &   2.61 & -2.44 &-31.43 &  7.59 &  7.69 &  7.63 &  7.60 &  7.60 & 3.0& BPS82b \\
     5778.45 & \eu{b}{3}{F}{}{3} - \eu{y}{3}{D}{\circ}{3} &   209  &   2.59 & -3.44 &-31.37 &  7.43 &  7.53 &  7.46 &  7.44 &  7.43 & 3.0& BKK91 \\
     4920.50 & \eu{z}{7}{F}{\circ}{5} - \eu{e}{7}{D}{}{4} &   318  &   2.83 &  0.07 &-30.51 &  7.43 &  7.48 &  7.45 &  7.43 &  7.43 & 2.5& OWL91   \\
     6229.23 & \eu{b}{3}{P}{}{1} - \eu{y}{3}{D}{\circ}{1} &   342  &   2.85 & -2.80 &-31.32 &  7.37 &  7.41 &  7.40 &  7.38 &  7.38 & 3.0& BKK91 \\
     6518.37 & \eu{b}{3}{P}{}{2} - \eu{y}{3}{D}{\circ}{3} &   342  &   2.83 & -2.46 &-31.37 &  7.45 &  7.55 &  7.48 &  7.46 &  7.45 & 3.0& BK94 \\
     5232.94 & \eu{z}{7}{P}{\circ}{4} - \eu{e}{7}{D}{}{5} &   383  &   2.94 & -0.06 &-30.54 &  7.40 &  7.48 &  7.42 &  7.40 &  7.40 & 2.2& OWL91   \\
     5281.79 & \eu{z}{7}{P}{\circ}{2} - \eu{e}{7}{D}{}{3} &   383  &   3.04 & -0.83 &-30.53 &  7.36 &  7.48 &  7.39 &  7.36 &  7.36 & 3.0& OWL91   \\
     4726.14 & \eu{z}{7}{P}{\circ}{3} - \eu{e}{5}{D}{}{2} &   384  &   3.00 & -3.25 &-30.34 &  7.63 &  7.72 &  7.65 &  7.63 &  7.63 & 3.3& FMW88 \\
     5807.78 & \eu{z}{5}{D}{\circ}{0} - \eu{e}{7}{D}{}{1} &   552  &   3.29 & -3.41 &-30.49 &  7.60 &  7.70 &  7.62 &  7.61 &  7.60 & 3.4& FMW88 \\
     5217.40 & \eu{z}{5}{D}{\circ}{4} - \eu{e}{5}{D}{}{3} &   553  &   3.21 & -1.07 &-30.37 &  7.45 &  7.53 &  7.47 &  7.45 &  7.45 & 2.9& BKK91 \\
     5324.18 & \eu{z}{5}{D}{\circ}{4} - \eu{e}{5}{D}{}{4} &   553  &   3.21 & -0.10 &-30.42 &  7.50 &  7.57 &  7.51 &  7.50 &  7.50 & 2.5& BKK91 \\
     5393.17 & \eu{z}{5}{D}{\circ}{3} - \eu{e}{5}{D}{}{4} &   553  &   3.24 & -0.72 &-30.42 &  7.46 &  7.53 &  7.48 &  7.46 &  7.46 & 2.9& BKK91 \\
     4808.15 & \eu{a}{3}{D}{}{3} - \eu{w}{3}{D}{\circ}{3} &   633  &   3.25 & -2.79 &-31.49 &  7.66 &  7.75 &  7.68 &  7.66 &  7.66 & 3.3& FMW88 \\
     5576.10 & \eu{z}{5}{F}{\circ}{1} - \eu{e}{5}{D}{}{0} &   686  &   3.43 & -1.00 &-30.32 &  7.58 &  7.66 &  7.60 &  7.59 &  7.59 & 3.0& FMW88 \\
     5586.76 & \eu{z}{5}{F}{\circ}{4} - \eu{e}{5}{D}{}{3} &   686  &   3.37 & -0.10 &-30.38 &  7.46 &  7.53 &  7.46 &  7.46 &  7.46 & 3.5& BKK91 \\
     6411.65 & \eu{z}{5}{P}{\circ}{2} - \eu{e}{5}{D}{}{3} &   816  &   3.65 & -0.60 &-30.38 &  7.46 &  7.54 &  7.48 &  7.46 &  7.46 & 3.1& BKK91 \\
     5397.62 & \eu{a}{1}{I}{}{6} - \eu{x}{3}{G}{\circ}{5} &   841  &   3.63 & -2.48 &-31.82 &  7.57 &  7.66 &  7.59 &  7.57 &  7.57 & 3.0& FMW88 \\
     5242.50 & \eu{a}{1}{I}{}{6} - \eu{z}{1}{H}{\circ}{5} &   843  &   3.63 & -0.97 &-31.56 &  7.58 &  7.61 &  7.60 &  7.58 &  7.58 & 3.2& OWL91   \\
     5379.58 & \eu{b}{1}{G}{}{4} - \eu{z}{1}{H}{\circ}{5} &   928  &   3.69 & -1.51 &-31.56 &  7.57 &  7.62 &  7.59 &  7.57 &  7.57 & 3.3& OWL91  \\
     5491.83 & \eu{c}{3}{F}{}{2} - \eu{u}{3}{D}{\circ}{3} &  1031  &   4.19 & -2.19 &-31.33 &  7.48 &  7.56 &  7.51 &  7.48 &  7.48 & 3.8& BK94 \\
     5236.20 & \eu{c}{3}{F}{}{2} - \eu{t}{3}{D}{\circ}{1} &  1034  &   4.19 & -1.50 &-31.32 &  7.39 &  7.47 &  7.41 &  7.39 &  7.39 & 3.2& OWL91  \\
     5607.66 & \eu{y}{5}{D}{\circ}{3} - \eu{e}{7}{G}{}{4} &  1058  &   4.15 & -2.27 &-30.36 &  7.57 &  7.67 &  7.60 &  7.58 &  7.57 & 3.0& FMW88 \\
     5858.78 & \eu{y}{5}{F}{\circ}{4} - \eu{f}{5}{F}{}{5} &  1084  &   4.22 & -2.26 &-30.40 &  7.57 &  7.66 &  7.60 &  7.58 &  7.57 & 3.3& FMW88 \\
     5638.26 & \eu{y}{5}{F}{\circ}{4} - \eu{g}{5}{D}{}{3} &  1087  &   4.22 & -0.87 &-30.52 &  7.64 &  7.70 &  7.67 &  7.65 &  7.64 & 3.0& FMW88 \\
     5662.52 & \eu{y}{5}{F}{\circ}{5} - \eu{g}{5}{D}{}{4} &  1087  &   4.18 & -0.57 &-30.52 &  7.58 &  7.63 &  7.60 &  7.59 &  7.59 & 3.0& OWL91   \\
     5197.94 & \eu{y}{5}{F}{\circ}{1} - \eu{f}{5}{P}{}{1} &  1091  &   4.30 & -1.64 &-30.25 &  7.65 &  7.74 &  7.68 &  7.66 &  7.66 & 3.0& FMW88 \\
     5522.45 & \eu{z}{3}{P}{\circ}{2} - \eu{g}{5}{D}{}{2} &  1108  &   4.21 & -1.55 &-30.46 &  7.64 &  7.71 &  7.67 &  7.64 &  7.64 & 3.2& FMW88 \\
     5517.06 & \eu{z}{3}{P}{\circ}{2} - \eu{e}{5}{P}{}{2} &  1109  &   4.21 & -2.37 &-30.21 &  7.85 &  7.95 &  7.88 &  7.86 &  7.85 & 3.3& FMW88 \\
     5295.31 & \eu{z}{5}{G}{\circ}{3} - \eu{e}{5}{H}{}{3} &  1146  &   4.42 & -1.69 &-30.16 &  7.66 &  7.73 &  7.69 &  7.66 &  7.66 & 3.2& FMW88 \\
     5367.47 & \eu{z}{5}{G}{\circ}{3} -  \eu{e}{5}{H}{}{4} &  1146  &   4.41 &  0.44 &-30.20 & 7.33 &  7.39 &  7.35 &  7.33 &  7.33 & 2.6& OWL91   \\
     5285.13 & \eu{z}{3}{G}{\circ}{4} - \eu{f}{3}{F}{}{4} &  1166  &   4.43 & -1.64 &-30.12 &  7.59 &  7.68 &  7.62 &  7.59 &  7.59 & 3.2& FMW88 \\
     6105.13 & \eu{y}{3}{F}{\circ}{4} - \eu{g}{5}{F}{}{5} &  1175  &   4.55 & -2.05 &-30.42 &  7.59 &  7.70 &  7.62 &  7.60 &  7.60 & 3.3& FMW88 \\
     5852.22 & \eu{y}{3}{F}{\circ}{4} - \eu{f}{5}{G}{}{4} &  1178  &   4.55 & -1.33 &-30.29 &  7.63 &  7.73 &  7.66 &  7.64 &  7.63 & 2.8& FMW88 \\
     5855.08 & \eu{y}{3}{F}{\circ}{3} - \eu{e}{5}{H}{}{4} &  1179  &   4.61 & -1.48 &-30.21 &  7.46 &  7.55 &  7.48 &  7.46 &  7.46 & 3.3& BK94 \\
     5930.18 & \eu{y}{3}{F}{\circ}{2} - \eu{e}{3}{G}{}{3} &  1180  &   4.65 & -0.23 &-30.19 &  7.54 &  7.60 &  7.57 &  7.54 &  7.54 & 2.8& FMW88 \\
     5679.02 & \eu{y}{3}{F}{\circ}{2} - \eu{f}{3}{F}{}{3} &  1183  &   4.65 & -0.92 &-30.07 &  7.74 &  7.83 &  7.78 &  7.75 &  7.75 & 3.1& FMW88 \\
\hline \noalign{\smallskip}
mean         &  &    &      &       &       & 7.53 & 7.62 & 7.56 & 7.54 & 7.54 & & \\
             &  &    &      &       &       & \scriptsize{$\pm$0.10} & \scriptsize{$\pm$0.11} & \scriptsize{$\pm$0.10} & \scriptsize{$\pm$0.10} & \scriptsize{$\pm$0.10} & & \\
\hline \\
 \multicolumn{13}{c}{ \ion{Fe}{ii} lines} \\
     4491.40 & \eu{b}{4}{F}{}{3/2} - \eu{z}{4}{F}{\circ}{3/2} & 37 & 2.84 & -2.76 &-32.02 &  7.61 &  7.60 &  7.60 &  7.61 &  7.61 & 3.5 &    RU98 \\
     4582.83 & \eu{b}{4}{F}{}{5/2} - \eu{z}{4}{F}{\circ}{7/2} & 37 & 2.83 & -3.22 &-32.03 &  7.54 &  7.52 &  7.54 &  7.54 &  7.54 & 3.4 &    RU98 \\
     4508.29 & \eu{b}{4}{F}{}{3/2} - \eu{z}{4}{D}{\circ}{1/2} & 38 & 2.84 & -2.34 &-32.00 &  7.46 &  7.44 &  7.46 &  7.46 &  7.46 & 3.5 &    RU98 \\
     4620.52 & \eu{b}{4}{F}{}{7/2} - \eu{z}{4}{D}{\circ}{7/2} & 38 & 2.82 & -3.29 &-32.02 &  7.48 &  7.47 &  7.48 &  7.48 &  7.48 & 3.5 &    RU98 \\
     6369.46 & \eu{a}{6}{S}{}{5/2} - \eu{z}{6}{D}{\circ}{3/2} & 40 & 2.89 & -4.25 &-32.06 &  7.59 &  7.59 &  7.59 &  7.59 &  7.59 & 3.7 &    RU98 \\
     6432.68 & \eu{a}{6}{S}{}{5/2} - \eu{z}{6}{D}{\circ}{5/2} & 40 & 2.89 & -3.71 &-32.07 &  7.62 &  7.61 &  7.62 &  7.62 &  7.62 & 3.6 &    RU98 \\
     5284.11 & \eu{a}{6}{S}{}{5/2} - \eu{z}{6}{F}{\circ}{7/2} & 41 & 2.88 & -3.13 &-32.04 &  7.47 &  7.46 &  7.47 &  7.47 &  7.47 & 3.5 &    M83 \\
     4923.93 & \eu{a}{6}{S}{}{5/2} - \eu{z}{6}{P}{\circ}{3/2} & 42 & 2.88 & -1.42 &-32.03 &  7.57 &  7.56 &  7.57 &  7.57 &  7.57 & 3.5 &    M83 \\
     5018.44 & \eu{a}{6}{S}{}{5/2} - \eu{z}{6}{P}{\circ}{5/2} & 42 & 2.88 & -1.23 &-32.04 &  7.56 &  7.54 &  7.56 &  7.56 &  7.56 & 3.5 &    M83 \\
     5991.38 & \eu{a}{4}{G}{}{11/2} - \eu{z}{6}{F}{\circ}{9/2} & 46 & 3.15 & -3.66 &-32.05 & 7.57 &  7.57 &  7.57 &  7.57 &  7.57 & 3.5 &    RU98 \\
     5264.81 & \eu{a}{4}{G}{}{5/2} - \eu{z}{4}{D}{\circ}{3/2} & 48 & 3.22 & -3.13 &-32.01 &  7.55 &  7.54 &  7.55 &  7.55 &  7.55 & 3.5 &    M83 \\
     5414.07 & \eu{a}{4}{G}{}{7/2} - \eu{z}{4}{D}{\circ}{7/2} & 48 & 3.21 & -3.65 &-32.02 &  7.56 &  7.56 &  7.57 &  7.57 &  7.57 & 3.8 &    RU98 \\
     5197.58 & \eu{a}{4}{G}{}{5/2} - \eu{z}{4}{F}{\circ}{3/2} & 49 & 3.22 & -2.34 &-32.02 &  7.55 &  7.53 &  7.55 &  7.55 &  7.55 & 3.4 &    RU98 \\
     5325.55 & \eu{a}{4}{G}{}{7/2} - \eu{z}{4}{F}{\circ}{7/2} & 49 & 3.21 & -3.32 &-32.03 &  7.64 &  7.62 &  7.64 &  7.64 &  7.64 & 3.6 &    RU98 \\
     5425.26 & \eu{a}{4}{G}{}{9/2} - \eu{z}{4}{F}{\circ}{9/2} & 49 & 3.20 & -3.38 &-32.04 &  7.62 &  7.60 &  7.62 &  7.62 &  7.62 & 3.5 &    RU98 \\
     6239.95 & \eu{b}{4}{D}{}{1/2} - \eu{z}{4}{P}{\circ}{3/2} & 74 & 3.89 & -3.57 &-32.00 &  7.59 &  7.59 &  7.59 &  7.59 &  7.59 & 3.7 &    RU98 \\
     6247.56 & \eu{b}{4}{D}{}{5/2} - \eu{z}{4}{P}{\circ}{3/2} & 74 & 3.89 & -2.43 &-32.00 &  7.59 &  7.57 &  7.59 &  7.59 &  7.59 & 3.4 &    RU98 \\
     6456.38 & \eu{b}{4}{D}{}{7/2} - \eu{z}{4}{P}{\circ}{5/2} & 74 & 3.90 & -2.18 &-32.00 &  7.60 &  7.58 &  7.60 &  7.60 &  7.60 & 3.6 &    RU98 \\
\hline \noalign{\smallskip}
mean         &  &    &      &       &       & 7.56 & 7.55 & 7.56 & 7.56 & 7.56 & & \\
             &  &    &      &       &       & \scriptsize{$\pm$0.05}& \scriptsize{$\pm$0.05}&\scriptsize{$\pm$0.05}&\scriptsize{$\pm$0.05}&\scriptsize{$\pm$0.05}& & \\
\hline  \noalign{\smallskip}
\multicolumn{13}{l}{$ ^1$ References to the adopted $gf-$values: } \\
\multicolumn{13}{l}{BKK91 = \citet{fe-BKK91}, BK94 = \citet{fe-BK94}, BIP79 = \citet{BIP79}, } \\
\multicolumn{13}{l}{BPS82a = \citet{BPS82a}, BPS82b = \citet{BPS82b}, FMW88 = \citet{FMW88}, } \\
\multicolumn{13}{l}{M83 = \citet{Moity83}, OWL91 = \citet{fe-OWL91}, RU98 = \citet{RU98} } \\
\end{longtable}
}

\end{document}